\documentclass[12pt,oneside]{article}
\usepackage{amsmath,amssymb,graphics}
\usepackage[left=3.2cm,top=3.6cm,bottom=3.6cm,right=3.2cm,head=0cm,foot=0.7cm]{geometry}
\newcommand{\field}[1]{\mathbb{#1}}

\usepackage[font=small]{caption}

\title{When Does a Boltzmannian Equilibrium Exist?}
\author{Charlotte Werndl\footnote{Department of Philosophy, University of Salzburg and
Department of Philosophy, Logic and Scienic Method, London School of Economics and Political Science,
charlotte.werndl@sbg.ac.at.}  \hspace{1mm}and Roman Frigg\footnote{Department of Philosophy, Logic and Scientic Method, and
Centre for Philosophy of Natural and Social Science, London School of Economics and Political Science. r.p.frigg@lse.ac.uk.}}

\date{\small{\vspace{5mm} To be published in:\\ Daniel Bedingham, Owen Maroney and Christopher Timpson (eds.): {\it Quantum Foundations of Statistical Mechanics}, Oxford: Oxford University Press.}}

\begin{document}
\maketitle

\tableofcontents

\section{Introduction}\label{Intro}

The received wisdom in statistical mechanics (SM) is that isolated systems, when left to themselves, approach equilibrium. But under what circumstances does an equilibrium state exist and an approach to equilibrium take place? In this paper we address these questions from the vantage point of the long-run fraction of time definition of Boltzmannian equilibrium that we developed in two recent papers (Werndl and Frigg 2015a, 2015b). After a short summary of Boltzmannian statistical mechanics (BSM) and our definition of equilibrium (Section \ref{BSM}), we state an existence theorem which provides general criteria for the existence of an equilibrium state (Section \ref{Theorem}). We first illustrate how the theorem works with a toy example (Section \ref{Simple Example}), which allows us to illustrate the various elements of the theorem in a simple setting. After commenting on the ergodic programme (Section \ref{Ergodic}) we discuss equilibria in a number of different gas systems: the ideal gas, the dilute gas, the Kac gas, the stadium gas, the mushroom gas and the multi-mushroom gas (Section \ref{Gases}). In the conclusion we briefly summarise the main points and highlight open questions (Section \ref{Conclusion}).

\section{Boltzmannian Equilibrium}\label{BSM}

Our focus are systems which, at the micro level, are measure-preserving deterministic dynamical systems $(X,\Sigma_{X},\mu_{X}, T_{t})$.\footnote{This section is based on Werndl and Frigg (2015a, 2015b). For a discussion of stochastic systems see Werndl and Frigg (2016).} The set $X$ represents all possible micro-states; $\Sigma_{X}$ is a $\sigma$-algebra of  subsets of $X$; the evolution function $T_{t}:X\rightarrow X$, $t\in\field{R}$ (continuous time) or $\field{Z}$ (discrete time) is a measurable function in $(t,x)$ such that $T_{t_{1}+t_{2}}(x)=T_{t_{2}}(T_{t_{1}}(x))$ for all $x\in X$ and all $t_{1},t_{2}\in\field{R}$ or $\field{Z}$; $\mu_{X}$ is a measure on $\Sigma_{X}$, which is invariant under the dynamics. That is, $\mu_{X}(T_{t}(A))=\mu_{X}(A)$ for all $A\in\Sigma_{X}$ and all $t$.\footnote{At this point the measure need not be normalised.} The function $s_{x}:\field{R}\rightarrow X$ or $s_{x}:\field{Z}\rightarrow X$, $s_{x}(t)=T_{t}(x)$ is called
the solution through the point $x\in X$.\\

A set of macro-variables $\{v_{1}, ..., v_{l}\}$ ($l \in \field{N}$) characterises the system at the macro-level. The fundamental posit of BSM is that macro-states supervene on micro-states, implying that a system's micro-state uniquely determines its macro-state. Thus the macro-variables are measurable functions $v_{i}:X\rightarrow \field{V}_{i}$, associating a value with a point $x\in X$. Capital letters $V_{i}$ will be used to denote the values of $v_{i}$. A particular set of values $\{V_{1}, ..., V_{l}\}$ defines a \emph{macro-state} $M_{V_{1}, \ldots, V_{l}}$. If the specific values $V_{i}$ do not matter, we only write `$M$' rather than `$M_{V_{1}, ..., V_{l}}$'. For now all we need is the general definition of macro-variables. We will discuss them in more detail in Sections~\ref{Theorem} and \ref{Simple Example}, where we will see that the choice of macro-variables is a subtle and important matter and that the nature as well as the existence of an equilibrium state crucially depends on it.\\

The determination relation between micro-states and macro-states will nearly always be many-to-one. Therefore, every macro-state $M$ is associated with a macro-region consisting of all micro-states for which the system is in $M$. A neglected but important issue is on what space macro-regions are defined. The obvious option would be $X$, but in many cases this is not what happens. In fact, often macro-regions are defined on a subspace $Z \subset X$. Intuitively speaking, $Z$ is a subset whose states evolve into the same equilibrium macro-state. To give an example: for a dilute gas with $N$ particles $X$ is the $6N$-dimensional space of all position and momenta, but $Z$ is the $6N-1$ dimensional energy hypersurface. $X$ will be called the \textit{full state space} and $Z$ the \textit{effective state space} of the system. The macro-region $Z_{M}$ corresponding to macro-state $M$ relative to $Z$ is then defined as the set of all $x \in Z$ for which  $M$ supervenes on $x$. A set of macro-states is complete relative to $Z$ iff (if and only if) it contains all states of $Z$. The members of a complete set of macro-regions $Z_{M}$ do not overlap and jointly cover $Z$, i.e. they form a partition of $Z$.\\

$Z$ has to be determined on a case-by-case basis, because the particulars of the system under consideration determine the correct choice of $Z$. We return to this point in Section~\ref{Theorem}. However, there is one general constraint on such a choice that needs to be mentioned now. Since a system can never leave the partition of macro-regions, it is clear that $Z$ must be mapped onto itself under $T_{t}$. If such a $Z$ is found, the sigma algebra on $X$ can be restricted to $Z$ and one can consider a measure on $Z$ which is invariant under the dynamics and normalized (i.e. $\mu_{Z}(Z)=1$). In this way the  measure-preserving dynamical system $(Z,\Sigma_{Z},\mu_{Z},T_{t})$  with a normalized measure $\mu_{Z}$ is obtained.\footnote{The dynamics is given by the evolution equations restricted to $Z$. We follow the literature by denoting it again by $T_{t}$.}  We call $(Z,\Sigma_{Z},\mu_{Z},T_{t})$ the \emph{effective system} (as opposed to the \emph{full system} $(X,\Sigma_{X},\mu_{Z},T_{t})$).\\

$M_{eq}$ is the equilibrium macrostate and the corresponding macro-region is $Z_{M_{eq}}$. An important aspect of the standard presentation of BSM is that $Z_{M_{eq}}$ is the largest macro-region. The notion of the `largest macro-region' can be interpreted in two ways. First, `largest' can mean that the equilibrium macro-region takes up a large part of $Z$. More specifically, $Z_{M_{eq}}$ is said to be \emph{$\beta$-dominant}
iff $\mu_{Z}(Z_{M_{eq}}) \geq\beta$ for a particular $\beta\in (\frac{1}{2},1]$. If $Z_{M_{eq}}$ is $\beta$-dominant, it is clear that it is also $\beta'$-dominant for all $\beta'$ in $(1/2, \, \beta)$. Second, `largest' can mean `larger than any other macro-region'. We say that $Z_{M_{eq}}$ is \emph{$\delta$-prevalent} iff $\min_{M \neq M_{eq}} [\mu_{Z}(Z_{M_{eq}}) -\mu_{Z}(Z_{M})]\geq\delta$ for a particular $\delta > 0$, $\delta\in\field{R}$. It follows that if a $Z_{M_{eq}}$ is $\delta$-prevalent, then it is also $\delta'$-prevalent for all  $\delta'$ in $(0, \, \delta)$. We do not adjudicate between these different definitions; either meaning of `large' can be used to define equilibrium. However, we would like to point out that they are not equivalent: if an equilibrium macro-region is $\beta$-dominant, there is a range of values for $\delta$ so that the macro-region is also $\delta$-prevalent for these values. However the converse fails.\\

Now the question is: why is the equilibrium state $\beta$-dominant or $\delta$-prevalent? A justification ought to be as close as possible to the thermodynamics (TD) notion of equilibrium. In TD a system is in equilibrium just in case change has come to a halt and all thermodynamic variables assume constant values (\textit{cf.} Reiss 1996, 3). This would suggest a definition of equilibrium according to which every initial condition lies on trajectory for which $\{v_{1}, ..., v_{k}\}$ eventually assume constant values. Yet this is unattainable for two reasons. First, because of Poincar\'{e} recurrence, the values of the $v_{i}$ will never reach a constant value and keep fluctuating. Second, in dynamical systems we cannot expect \textit{all} initial conditions to lie on trajectories that approach equilibrium (see, e.g., Callender 2001).\\

To do justice to these facts about dynamical systems we revise the TD definition slightly and define equilibrium as the macro-state in which trajectories starting in most initial conditions spend most of their time. This is not a feeble compromise. Experimental results show that physical systems exhibit fluctuations away from equilibrium (Wang et al.\ 2002). Hence strict TD equilibrium is actually unphysical and a definition of equilibrium that makes room for fluctuations is empirically more adequate.\\

To make this idea precise we introduce the long-run fraction of time a system spends in a region $A\in \Sigma_{Z}$ when the system starts in micro-state $x$ at time $t=0$:
\begin{eqnarray}\label{LF}
LF_{A}(x)&=&\lim_{t\rightarrow\infty}\frac{1}{t}\int_{0}^{t}1_{A}(T_{\tau}(x))d\tau\,\,\textnormal{for continuous time, i.e.}\,\,t\in\field{R},\,\,\\
LF_{A}(x)&=&\lim_{t\rightarrow\infty}\frac{1}{t}\sum_{\tau=0}^{t-1}1_{A}(T_{\tau}(x))\,\,\textnormal{for discrete time, i.e.}\,\,t\in\field{Z},\nonumber
\end{eqnarray}
\noindent where $1_{A}(x)$ is the characteristic function of $A$, i.e.\ $1_{A}(x)=1$ for $x\in A$ and $0$ otherwise. Note that a measure-preserving dynamical system $(Z, \Sigma_{Z},\mu_{Z},T_{t})$ with the normalized measure $\mu_{Z}$ is \emph{ergodic} iff for any $A \in \Sigma_{Z}$:
\begin{equation}\label{ergodicE}
LF_{A}(x)=\mu_{Z}(A),
\end{equation}
for all $x \in Z$ except for a set $W$ with $\mu_{Z}(W)=0$.\\

The locution `most of their time' is beset with the same ambiguity as the `largest macro-state'. On the first reading `most of the time' means more than half of the total time. This leads to the following formal definition of equilibrium:

\begin{quote}
\textit{BSM $\alpha$-$\varepsilon$-Equilibrium.} Consider an isolated system $S$ whose macro-states are specified in terms of the macro-variables $\{v_{1}, ..., v_{k}\}$ and which, at the micro level, is a measure-preserving deterministic dynamical system $(Z,\Sigma_{Z},\mu_{Z},T_{t})$. Let $\alpha$ be a real number in $(0.5, 1]$, and let $1 \gg \varepsilon \ge 0$ be a very small real number. If there is a macrostate $M_{V_{1}^{*}, ..., V_{k}^{*}}$ satisfying the following condition, then it is the $\alpha$-$\varepsilon$-equilibrium state of $S$: There exists a set $Y\subseteq Z$ such that $\mu_{Z}(Y)\geq 1-\varepsilon$, and all initial states $x\in Y$ satisfy
\begin{equation}\label{alpha}
LF_{Z_{M_{V_{1}^{*}, ..., V_{l}^{*}}}}\!(x) \, \geq \, \alpha.
\end{equation}
We then write $M_{\alpha\textnormal{-}\varepsilon\textnormal{-}eq}\, := \, M_{V_{1}^{*}, ..., V_{k}^{*}}$.
\end{quote}
\noindent An obvious question concerns the value of $\alpha$. Often the assumption seems to be that $\alpha$ is close to one. This is reasonable but not the only possible choice. For our purposes nothing hangs on a the value of $\alpha$ and so we leave it open what the best choice would be.\\

On the second reading `most of the time' means that the system spends more time in the equilibrium macro-state than in any other macro-state. This idea can be rendered precise as follows:
\begin{quote}
\textit{BSM $\gamma$-$\varepsilon$-Equilibrium.}
Consider an isolated system $S$ whose macro-states are specified in terms of the macro-variables $\{v_{1}, ..., v_{k}\}$ and which, at the micro level, is a measure-preserving deterministic dynamical system $(Z,\Sigma_{Z},\mu_{Z}, T_{t})$. Let $\gamma$ be a real number in $(0, 1]$ and let $1 \gg \varepsilon \geq 0$ be a very small real number. If there is a macro-state $M_{V_{1}^{*}, ..., V_{l}^{*}}$ satisfying the following condition, then it is the $\gamma$-$\varepsilon$ equilibrium state of $S$: There exists a set $Y\subseteq Z$ such that $\mu_{Z}(Y)\geq 1-\varepsilon$ and for all initial conditions $x\in Y$:
\begin{equation}\label{gamma}
LF_{Z_{M_{V_{1}^{*}, ..., V_{l}^{*}}}}\!(x) \, \geq \,
LF_{Z_{M}}\!(x)+\gamma
\end{equation}
\noindent for all macro-states $M \neq M_{V_{1}^{*}, ..., V_{l}^{*}}$. We then write $M_{\gamma\textnormal{-}\varepsilon\textnormal{-}eq}\, := \, M_{V_{1}^{*}, ..., V_{k}^{*}}$.
\end{quote}

\noindent As above, nothing in what we say about equilibrium depends on the particular value of the parameter $\gamma$ and so we leave it open what the best choice would be.\\

We contend that these two definitions provide the relevant notion of equilibrium in BSM. It is important to emphasise, that they remain silent about the size of equilibrium macro-regions, and do not in any obvious sense imply anything about seize. Indeed, equilibrium marco-regions being extremely small would be entirely compatible with the definitions. That these macro-regions have the right size is a result established in the following two theorems:

\begin{quote}
\emph{Dominance Theorem}: If  $M_{\alpha\textnormal{-}\varepsilon\textnormal{-}eq}$ is an $\alpha$-$\varepsilon$-equilibrium of system $S$, then $\mu_{Z}(Z_{M_{\alpha\textnormal{-}\varepsilon\textnormal{-}eq}}) \geq \beta$ for $\beta=\alpha(1-\varepsilon)$.\footnote{We assume that $\varepsilon$ is small enough so that $\alpha(1-\varepsilon)> \frac{1}{2}$.}
\end{quote}

\begin{quote}
\emph{Prevalence Theorem}: If $M_{\gamma\textnormal{-}\varepsilon\textnormal{-}eq}$ is a $\gamma$-$\varepsilon$-equilibrium of system $S$, then
$\mu_{Z}(Z_{M_{\gamma\textnormal{-}\varepsilon\textnormal{-}eq}}) \geq \mu_{Z}(Z_{M})+ \delta$ for $\delta = \gamma-\varepsilon$ for all macro-states $M\neq M_{\gamma\textnormal{-}\varepsilon\textnormal{-}eq}$.\footnote{We assume that $\varepsilon<\gamma$.}
\end{quote}

\noindent Both theorems are completely general in that no dynamical assumptions are made (in particular it is not assumed that systems are ergodic -- \emph{cf.}\ Equation \ref{ergodicE}), and hence the theorems also apply to strongly interacting systems.\\

An important aspect of the above definitions of equilibrium that the presence of an approach to equilibrium is built into the notion of an equilibrium state. If a state is not such that the system spends most of the time in that state (in one of the two senses specified), then that state simply is not an equilibrium state. In other words, if the system does not approach equilibrium, then there is no equilibrium. Having an equilibrium state and there being an approach to equilibrium are two sides of the same coin.\\

The theorems make the conditional claim that \textit{if} an equilibrium exits, \textit{then} it is large in the relevant sense. Some systems do not have equilibria. If, for instance, the dynamics is given by the identity function, then no approach to equilibrium takes place, and the antecedent of the conditional is wrong. As with all conditionals, the crucial question is whether, and under what conditions, the antecedent holds. We turn to this issue now. As we have just seen, the question whether there is an equilibrium state is tantamount to the question whether the approach to equilibrium takes place, and so the issue of existence is not merely an inconsequential subtlety in mathematical physics - it concerns one of the core questions in SM.

\section{The Existence of an Equilibrium Macro-State}\label{Theorem}

We now turn to the core question of this paper: under what circumstances does a Boltzmannian equilibrium macro-state exist? The main message is that for an equilibrium to exist three factors need to cooperate: the choice of macro-variables, the dynamics of the system, and the choice of the effective state space $Z$. The cooperation between these factors can take different forms and there is more than one constellation that can lead to the existence of an equilibrium state. The important point is that the answer to the question of existence is holistic: it not only depends on three factors rather than one, but also on the interplay between these factors. For these reasons we call these three factors the \textit{holist trinity}.\\

A number of previous proposals fail to appreciate this point. The problem of the approach to equilibrium has often been framed as the challenge to identify \textit{one} crucial property and show that the relevant systems possess this property. We first introduce the trinity in an informal way and illustrate it with examples, showing what requisite collaborations look like and what can go wrong. This informal presentation is followed by a rigorous mathematical theorem providing necessary and sufficient conditions for the existence of an equilibrium state.

\subsection{The Holist Trinity}\label{HT}
\textit{\textbf{Macro-variables}.} The first condition is that the macro-variables must be the `right' ones: the same system can have an equilibrium with respect to one set of macro-variables and fail to have an equilibrium with respect to another set of macro-variables. The existence of an equilibrium depends as much on the choice of macro-variables as it depends on the system's dynamical properties. Different choices are possible, and these choices lead to different conclusions about the equilibrium behaviour of the system. This will be illustrated below in Section \ref{Simple Example} with the example of the simple pendulum.\footnote{For further examples see Werndl and Frigg (2015a).}\\

This also implies that if no macro-variables are introduced, considerations of equilibrium make no sense at all. Obvious as this may seem, some confusion has resulted from ignoring this simple truism. Sklar (1973, 209) mounts an argument against the ergodic approach by pointing out that a system of two hard spheres in a box has the right dynamics (namely ergodicity) and yet fails to show an approach to equilibrium. It hardly comes as a surprise, though, that there is no approach to equilibrium if the system has no macro-variables associated with it in terms of which equilibrium could even be defined.\\

\noindent\textit{\textbf{Dynamics}.} The existence of an equilibrium depends as much on the dynamics of the system as it depends on the choice of macro-variables. Whatever the macro-variables, if the dynamics does not `collaborate', then there is no approach to equilibrium. For this reason the converses of the Dominance and Prevalence Theorems fail: it is \textit{not} the case that if there is a $\beta$-dominant/$\delta$-prevalent macro-region, then this macro-region corresponds to a $\alpha$-$\varepsilon$-equilibrium/$\gamma$-$\varepsilon$-equilibrium. If, for instance, the dynamics is the identity function, then there can be no approach to equilibrium because states in a small macro-region will always stay in this region. Or assume that there is system whose dynamics is such that micro-states that are initially in the largest macro-region always remain in the largest macro-region and states initially in smaller macro-regions only evolve into states in these smaller macro-regions. Then there is no approach to equilibrium because non-equilibrium states will not evolve into equilibrium. This point will also be illustrated with the example of the simple pendulum in Section \ref{Simple Example}.\\

\noindent \textit{\textbf{Identifying $Z$}.}
A number of considerations in connection with equilibrium depend on the choice of the effective state space $Z$, which is the set relative to which macro-regions are defined. Indeed, the existence of an equilibrium state depends on the correct choice of $Z$. There can be situations where a system has an equilibrium with respect to one choice of $Z$ but not with respect another choice of $Z$. One can choose $Z$ too small, and, as a consequence, it will not be true that most initial states approach equilibrium and hence there will be no equilibrium (recall that on our definition the system has no equilibrium if it does't approach equilibrium). One can, however, make the opposite mistake and choose $Z$ too large. If there is an equilibrium relative to some set $Z$, it need not be the case that an equilibrium exists also on a superset of this set. So $Z$ can be chosen too large as well as too small. That $Z$ can be chosen too large will be illustrated with the example of the simple pendulum in Section \ref{Simple Example}.\\

There is no algorithmic procedure to determine $Z$, but one can pinpoint a number of relevant factors. The most obvious factors are constraints and boundary conditions imposed on the system. If a system cannot access certain parts of $X$, then these parts are not in $Z$. In all examples below we see parts of $X$ being `cut off' when constructing $Z$ because of mechanical restrictions preventing the system from entering certain regions. Another important factor in determining $Z$ are conserved quantities. Their role, however, is less clear-cut than one might have hoped for. It is not universally true that $Z$ has to lie within a hyper-surface of conserved quantities. Whether $Z$ is so constrained depends on the macro-variables. Consider the example of energy. In some cases (the dilute gas in Section \ref{Gases}, for instance), equilibrium values depend on the energy of the system (equilibrium states are different for different energies) and hence $Z$ must lie within an energy hyper-surface. In other cases (the oscillator in Section \ref{Simple Example}, for instance) equilibrium is insensitive toward changes in the system's energy (the equilibrium state is the same for all energy values) and therefore $Z$ is not confined to an energy hyper-surface. This brings home again the holist character of the issue: $Z$ not only depends on mechanical invariants and constraints, but also on the macro-variables. \\

The interplay between these factors is illustrated with a simple toy model in Section \ref{Simple Example}. Due to its simplicity it is tangible how the three factors mutually constrain each other and it becomes clear how sensitively the existence of an equilibrium depends on the careful balance of these factors. In Section \ref{Gases} we  discuss how these considerations play out in different gas systems.

\subsection{The Existence Theorem }\label{Existence Theorem}

In this subsection we present the Equilibrium Existence Theorem, a theorem providing necessary and sufficient conditions for the existence of an equilibrium state (either of the $\alpha$-$\varepsilon$ or the $\gamma$-$\varepsilon$ type).\footnote{The proof is given in Werndl and Frigg (2015a).} Before stating the theorem we have to introduce another theorem, the Ergodic Decomposition Theorem (\emph{cf.} Petersen 1983, 81). An ergodic decomposition of a system is a partition of the state space into cells so that the cells are invariant under the dynamics (i.e., are mapped onto themselves) and that the dynamics within each cell is ergodic (\emph{cf.} equation \ref{ergodicE} for the definition of ergodicity).\footnote{It is allowed that the cells are of measure zero and that there are uncountably many of them.} \\

The Ergodic Decomposition Theorem says that such a decomposition exists for \textit{every} measure-preserving dynamical system with a normalised measure, and that the decomposition is unique. In other words, the dynamics of a system can be as complex as we like and the interactions between the constituents of the system can be as strong and intricate as we like, and yet there exists a unique ergodic decomposition of the state space of the system. A simple example of the theorem is the harmonic oscillator: the ellipses around the coordinate origin are the cells of the partition and the motion on the ellipses is ergodic.\\

For what follows it is helpful to have a more formal rendering of an ergodic decomposition. Consider the system $(Z,\Sigma_{Z},\mu_{Z},T_{t})$. Let $\Omega$ be an index set (which can but need not be countable), which comes equipped with a probability measure $\nu$. Let $Z_{\omega}$, $\omega \in \Omega$, be the cells into which the system's state space can be decomposed, and let $\Sigma_{\omega}$ and $\mu_{\omega}$, respectively, be the sigma algebra and measure defined on $Z_{\omega}$. These can be gathered together in  `components' $C_{\omega}=(Z_{\omega}, \Sigma_{\omega}, \mu_{\omega}, T_{t})$. The Ergodic Decomposition Theorem says that for every system $(Z,\Sigma_{Z},\mu_{Z},T_{t})$ there exists a unique set of ergodic $C_{\omega}$ so that the system itself amounts to the collection of all the $C_{\omega}$. How the ergodic decomposition theorem works will be illustrated with the example in the next section.\\

We are now in a position to state our core result:

\begin{quote}
\noindent \emph{\textbf{Equilibrium Existence Theorem}}:
Consider a measure-preserving system $(Z,\Sigma_{Z},\mu_{Z},T_{t})$ with macro-regions $Z_{M_{V_{1}, \ldots,V_{l}}}$ and let $C_{\omega}=(Z_{\omega},\Sigma_{\omega},\mu_{\omega},T_{t})$, $\omega \in \Omega$, be its ergodic decomposition. Then the following two biconditionals are true:\\\\
\noindent \textit{$\alpha$-$\varepsilon$-equilibrium}:
\noindent There exists an $\alpha$-$\varepsilon$-equilibrium iff there is a macro-state $\hat{M}$ such that for every $C_{\omega}$:
\begin{equation}\label{ggg}
\mu_{\omega}(Z_{\omega}\!\cap\! Z_{\hat{M}}) \, \geq \, \alpha,
\end{equation}
except for components $C_{\omega}$ with $\omega\in\Omega'$, $\mu_{Z}(\cup_{\omega\in\Omega'}Z_{\omega})\leq\varepsilon$. $\hat{M}$ is then the $\alpha$-$\varepsilon$-equilibrium state.\\\\
\noindent \textit{$\gamma$-$\varepsilon$-equilibrium}:
\noindent There exists a $\gamma$-$\varepsilon$-equilibrium iff there is a macro-state $\hat{M}$ such that for every $C_{\omega} $ and any $M \neq \hat{M}$
\begin{equation}\label{stress}
\mu_{\omega}(Z_{\omega}\!\cap\! Z_{\hat{M}}) \, \geq \, \mu_{\omega}(Z_{\omega}\!\cap\! Z_{M})+\gamma,
\end{equation}
except for components $C_{\omega}$ with $\omega\in\Omega'$, $\mu_{Z}(\cup_{\omega\in\Omega'}Z_{\omega})\leq\varepsilon$. $\hat{M}$ is then the $\gamma$-$\varepsilon$-equilibrium state.
\end{quote}

\noindent Like the theorems we have seen earlier, the Equilibrium Existence Theorem is fully general in that it makes no assumptions about the system's dynamics other than that it be measure-preserving. Intuitively the theorems say that there is an $\alpha$-$\varepsilon$-equilibrium ($\gamma$-$\varepsilon$-equilibrium) iff if the system's state space is split up into invariant regions on which the motion is ergodic and the equilibrium macro-state takes up at least $\alpha$ of each region (the equilibrium region is at least $\gamma$ larger than any other macro-region), except, possibly, for regions of total measure $\varepsilon$. If we have found a space that meets these conditions, then it plays the role of the effective state space $Z$.\\

It is important to note that there may be many different macro-state/dynamics/$Z$ triplets that make the Existence Theorem true. The Theorem gives the foundation for a research programme aiming to find and classify such triplets. But before discussing a number of interesting cases, we want to illustrate the theorem in the simplest possible setting. This is our task in the next section.

\section{Toy Example: The Ideal Pendulum}\label{Simple Example}
Consider an ideal pendulum: a small mass $m$ hanging on a $1$ meter long massless string from the ceiling. The mass moves without friction. When displaced, the mass will oscillate around its midpoint. We displace the pendulum only in one spatial direction and so the motion takes place in plane perpendicular to the ceiling. The weight of the bob $mg$, where $g$ is the gravitational constant, has components parallel and perpendicular to the rod. The component perpendicular to the rod is $-mg\sin(x)$, where $x$ is the angular displacement. This component accelerates the bob, and hence we can apply Newton's second law:
\begin{equation}
m\frac{d^{2}x}{dt^{2}}=-mg\sin(x)
\end{equation}
For the simple pendulum the further assumption is made that the angular displacement is small (of absolute value less than $15$ degrees). Then $\sin(x)\approx x$, and the equation reduces to:
\begin{equation}\label{SP}
\frac{d^{2}x}{dt^{2}}=-gx.
\end{equation}
This equation describes simple harmonic motion. \\

That is, the full phase space $X$ is given by the possible angular displacement and angular velocity coordinates $(x,v)$, where the angular displacement is assumed to be less than $15$ degrees; and thus the displacement as well as the velocity is bounded from above).  Solving the differential equation $(\ref{SP})$ above gives
\begin{eqnarray}\label{harmonic-ellipses}
x(t)&=&A\cos(\lambda t - \phi) \nonumber \\
v(t)&=&\frac{dx}{dt}=-A\lambda \sin(\lambda t - \phi),
\end{eqnarray}
where $\lambda=\sqrt{g}$, $A$ is the amplitude (the maximum displacement from the midpoint),   and $\phi$ is the phase (the shift of the cosine and sinus functions along the time axis). $A$ and $\phi$ are determined by the initial angular displacement and initial angular velocity.\\

From these equations we see that the solutions $T_{t}(x,v)$ are ellipses and the full phase space $X$ is composed of these ellipses. This is illustrated in Figure 1. $\Sigma_{X}$ is the Borel $\sigma$-algebra on $X$, and the measure $\mu_{X}$ on the phase space $X$ is the normalized uniform measure. Taking these elements together yields the measure-preserving dynamical system $(X, \Sigma_{X},\mu_{X},T_{t})$. \\

\begin{figure}
\centering
\resizebox{0.5\textwidth}{!}{\includegraphics{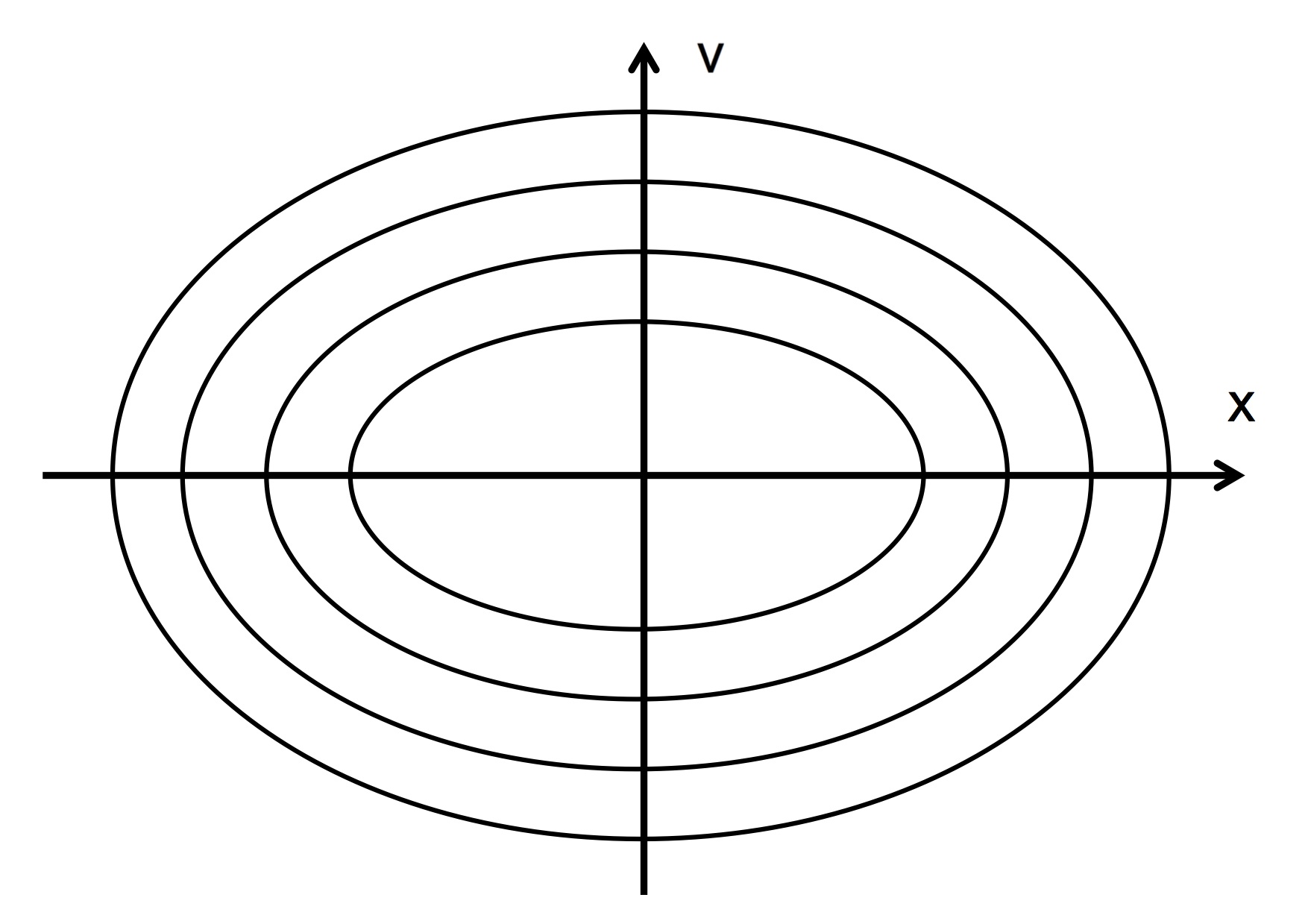}}\newline

\begin{center}
\small{Figure 1 --- The ergodic decomposition of the harmonic oscillator.}
\end{center}

\end{figure}

The effective phase space, i.e.\ the phase space relative to which equilibrium is defined, is in this case identical with $X$, and thus $(Z, \Sigma_{Z},\mu_{Z}, T_{t})=(X, \Sigma_{X},\mu_{X}, T_{t})$. We now illustrate the roles the macro-variables, the dynamics and the effective state space play in securing the existence of an equilibrium by discussing different choices and showing how they affect the existence of an equilibrium. \\

\noindent\textbf{The role of macro-variables.} \\

Consider $(Z,\Sigma_{Z}, \mu_{T}, T_{t})$ with the colour macro-variable $v_{c}$, a light bulb that can emit red and white light. So $\field{V}_{c}=\{r, w\}$, were `$r$' stands for red and `$w$' for white. The mapping is as follows: if the pendulum is on the right hand side of the midpoint \emph{and} on its way back to the midpoint, then the light is red; the light is white otherwise. This defines two macro-states $M_{r}$ and $M_{w}$. The macro-region $Z_{M_{r}}$ is the grey area in Figure 2 and $Z_{M_{w}}$ is the white area. \\

\begin{figure}
\centering
\resizebox{0.5\textwidth}{!}{\includegraphics{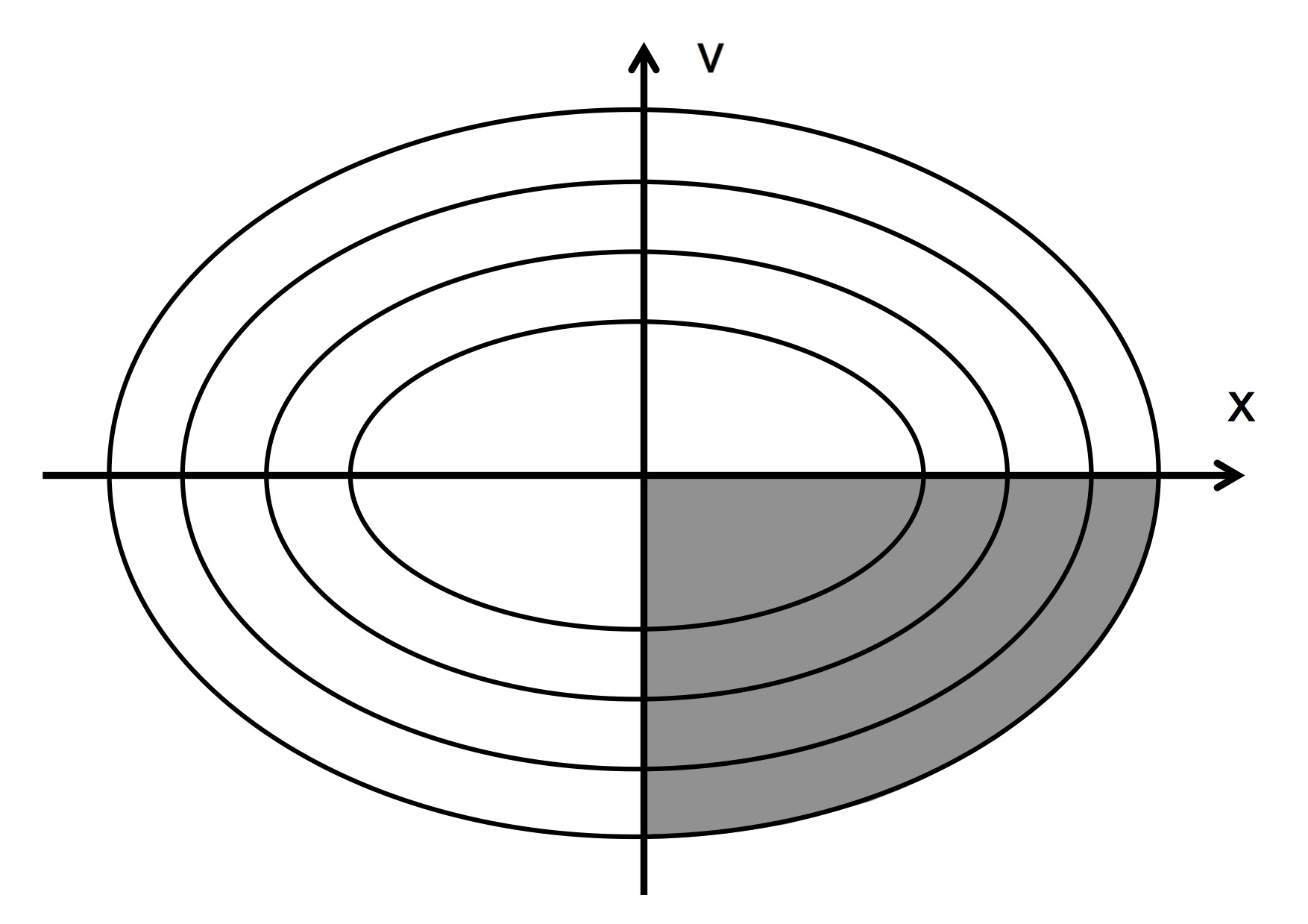}}\newline

\begin{center}
\small{Figure 2 --- The colour macro-variable $v_{c}$: if system's state is in the grey area the light is red; if it is in the white area the light is white.}
\end{center}
\end{figure}

Since the ideal pendulum oscillates with a constant frequency, $M_{w}$ is a $0.75$-$0$-equilibrium of the $\alpha$-$\varepsilon$ type: on each trajectory the light is white for three-quarters of the time and red for one quarter of the time. Thus, by the Dominance Theorem, $\mu(Z_{M_{w}}) \geq 0.75$ (and we have $\mu(Z_{M_{r}})=0.25$). $M_{w}$ is in fact also a  $0.5$-$0$-equilibrium of the $\gamma$-$\varepsilon$ type because the systems spends 0.5 more time in $M_{w}$ than in $M_{r}$. Thus by the Prevalence Theorem:  $\mu(Z_{M_{w}}) \geq \mu(Z_{M_{r}}) + 0.5$ for all $Z_{M_{r}}$.\\

Let us now discuss how the situation presents itself in terms of the Existence Theorem. To this end we first have a look at the ergodic decomposition theorem. The theorem says that $Z$ can be decomposed into components $C_{\omega}=(Z_{\omega}, \Sigma_{\omega}, \mu_{\omega}, T_{t})$. In the case of the harmonic oscillator the ergodic decomposition is the (uncountable) family of ellipses given by Equation \ref{harmonic-ellipses} and shown in Figure 1. Each $Z_{\omega}$ is a two-dimensional ellipse determined by the initial energy (the energy is determined by the initial displacement and velocity coordinates $(x, v)$). The $Z_{\omega}$ are the ellipses themselves; $\Sigma_{\omega}$ and $\mu_{\omega}$ are the standard Borel sets on a line and the normalised length measure on a line; and $T_{t}$ is the time evolution given by Equation \ref{harmonic-ellipses} restricted to the ellipses. It is easy to see that the motion on each ellipse is ergodic. The decomposition is parameterised by $\omega$, which in this case has a physical interpretation: it is the energy of the system. $\Omega$ is the (uncountable) set of energy values between zero and the energy corresponding to a 15 degrees angular displacement, and the measure $\nu$ on $\Omega$ is the standard Lebesgue measure. \\

Equation (\ref{ggg})  holds true for every component $C_{\omega}=(Z_{\omega},\Sigma_{\omega},\mu_{\omega}, T_{t})$ because on each ellipse three-quarters of the states correspond to a white light and one quarter to a red light, and hence $\mu_{\omega}(Z_{\omega}\!\cap\! Z_{M_{w}}) \, \geq \, 0.75$. Hence $M_{w}$ satisfies the condition for an $\alpha$-$\varepsilon$-equilibrium with $\alpha = 0.75$ and $\varepsilon = 0$. Likewise, equation (\ref{stress})  holds true for every component $C_{\omega}=(Z_{\omega},\Sigma_{\omega},\mu_{\omega}, T_{t})$ because on each ellipse three-quarters of the states correspond to a white light and one quarter to a red light, and hence $\mu_{\omega}(Z_{\omega}\!\cap\! Z_{M_{w}}) \, \geq \, \mu_{\omega}(Z_{\omega}\!\cap\! Z_{M})+0.5$ for all $Z_{M}\neq Z_{M_{w}}$. Hence $M_{w}$ satisfies the condition for an $\gamma$-$\varepsilon$-equilibrium with with $\gamma = 0.5$ and $\varepsilon = 0$. \\

Now consider a different macro-variable $v'_{c}$. It is defined like $v_{c}$ but with one crucial difference: the light is red when the pendulum is on the right side irrespective of whether it is moving towards or away from the midpoint. The light is white when the pendulum is on the left side or exactly in the middle. This is illustrated in Figure 3, where $Z_{M_{r}}$ is the grey and $Z_{M_{w}}$ is the white area. With respect to $v'_{c}$ the system has no equilibrium. For all solutions the red and the white light are each on half of the time, and both macrostates have equal measure $0.5$. \\

\begin{figure}
\centering
\resizebox{0.5\textwidth}{!}{\includegraphics{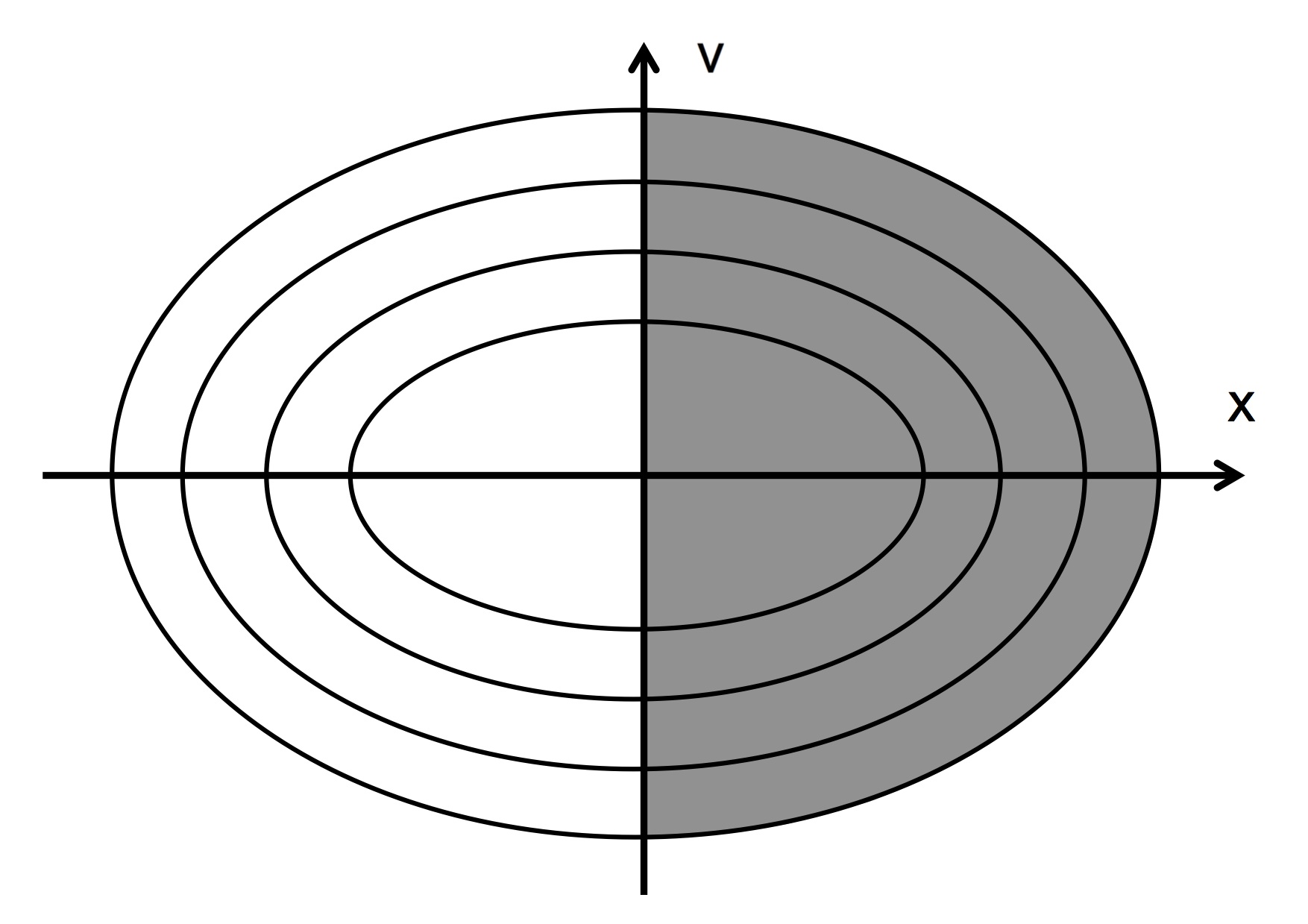}}\newline

\begin{center}
\small{Figure 3 --- The light bulb macro-variable $v'_{c}$: if system's state is in the grey area the light is red; if it is in the white area the light is white.}
\end{center}
\end{figure}

From the vantage point of the Existence theorem, the situation presents itself as follows. Equations (\ref{ggg}) and (\ref{stress}) cannot hold true any more because for every component $C_{\omega}=(Z_{\omega},\Sigma_{\omega},\mu_{\omega}, T_{t})$ half of the states correspond to a white light and a half of the states correspond to a red light. Hence the conditions of the Existence Theorem are not satisfied. This example illustrates that a small change in the macro-variable is enough to take us from a situation in which an equilibrium exists to one in which there is no equilibrium.\\

\noindent\textbf{The role of the dynamics}\\

As we have just seen, there exist equilibria of both types for the simple pendulum  $(Z,\Sigma_{Z},\mu_{Z}, T_{t})$ with the macro-variable $v_{c}$. We now change the dynamics: place a wall of negligible width exactly at the midpoint (perpendicular to the plane of motion) and assume that the pendulum bounces elastically off the wall. Denote this dynamics by $T'_{t}$. If the pendulum starts on the right hand side, it will always stay on the right hand side. On that side the white and the red light are on half of the time each and so the system has no equilibrium for initial conditions on the right hand side. This violates the condition (in both definitions of equilibrium) that there is at most a small set of initial conditions (of measure $ < \varepsilon$) for which the system does not satisfy the relevant equations (Equations \ref{alpha} and \ref{gamma} respectively). Hence, the system $(Z,\Sigma_{Z},\mu_{Z}, T'_{t})$ with the macro-variable $v_{c}$ has no equilibrium.\\

Let us look at the situation through the lens of the Existence Theorem. The ergodic decomposition is now more complicated than above. There are again uncountably many components $C'_{\omega}$. Yet because of the different dynamics, they are half-ellipses rather than ellipses. More specifically, the index set is $\Omega'=\Omega_{1}\cup\Omega_{2}$, where $\Omega_{1}$ consist of the uncountably different values of the energy for systems that start out on the right hand side, and $\Omega_{2}$ consist of the uncountably different values of the energy for systems that do not start on the right hand side. The measure on $\mu'_{\omega}$ on $\Omega'$ is defined by the condition that $\mu'_{\omega}$ restricted to both $\Omega_{1}$ and $\Omega_{2}$ is the Lebesgue measure divided by 2. Each $Z'_{\omega}$ is a two dimensional half-ellipse determined by the initial energy and whether the system starts on the right or the left. The sigma-algebra $\Sigma'_{\omega}$ is the usual Borel $\sigma$-algebra on $Z'_{\omega}$ and the measure $\mu'_{\omega}$ is the normalized length measure of the half-ellipse  $Z'_{\omega}$. The dynamics on the ellipses is again given by the restriction of $T'_{t}$ to the half-ellipses $Z'_{\omega}$. Taking these elements together gives us the components $C'_{\omega}=(Z'_{\omega},\Sigma'_{\omega},\mu'_{\omega}, T_{t}')$, and it is clear that the motion on each component is ergodic.\\

Now consider the components $C_{\omega}$ that correspond to the case where the pendulum starts on the right hand side. Note that measure $\mu_{Z}$ of all these components taken together is $1/2$. Yet half of any of these components is made up of states corresponding to the light being white and the remaining half is made up of states for which the light is red. Consequently, for these components equations (\ref{ggg}) and (\ref{stress}) cannot hold true. Thus the Existence Theorem is not satisfied because the condition is violated that there is at most a small set of initial conditions for which the system does not satisfy the relevant equations.\\

\noindent\textbf{The role of the effective phase space}\\

So far we discussed  a simple pendulum with a one-dimension position coordinate. Let us now consider the different setup where the pendulum's position coordinate is not one-dimensional but two-dimensional (and, again, we impose the constraint that the maximum displacement in any spatial direction is $\leq 15^{\circ}$), allowing the pendulum to oscillate in two directions, $x$ and $y$.
We now impose the constraint that the pendulum oscillates along a line going through the coordinate origin, and the time evolution $T''_{t}$  along this line is in fact the same as above, but now described with two-dimensional angular displacement coordinates. The full state space of the system $X''$ is thus a three-dimensional ellipsoid: the first two coordinates are the displacement coordinates  $x$ and $y$, and the third coordinate gives the velocity along the line cutting through the origin. $\Sigma_{X''}$ is again the Borel $\sigma$-algebra on $X''$, and $\mu_{X''}$  is the uniform measure on the ellipsoid. Then $(X'', \Sigma_{X''}, \mu_{X''}, T''_{t})$ is a measure-preserving dynamical system.\\

Now consider the two-dimensional colour macro variable $v''_{c}$, which can take three values: red, white and blue. So $\field{V''}_{c}=\{r, w, b\}$. Because the displacement coordinates are constrained to a line, the displacement coordinate of a solution either oscillates between the first and the third quadrant or between the second and the fourth quadrant. Suppose that if the pendulum is in the first quadrant, the light is red; if the pendulum is in the second quadrant, then the light is blue if the pendulum is on its way to back to the midpoint and white if it moves away from the midpoint or is exactly at the midpoint.  If the pendulum is in the third quadrant, then the light is red if the pendulum is on its way back to the midpoint and white if it moves away from the midpoint or is exactly at the midpoint. If the pendulum is in the fourth quadrant the light is white. This is illustrated in Figure 4. It is then easy to see that $\mu_{X''}(X''_{M_{w}})=1/2$, $\mu_{X''}(X''_{M_{r}})=3/8$ and $\mu_{X''}(X''_{M_{b}})=1/8$. \\

\begin{figure}
\centering
\resizebox{0.7\textwidth}{!}{\includegraphics{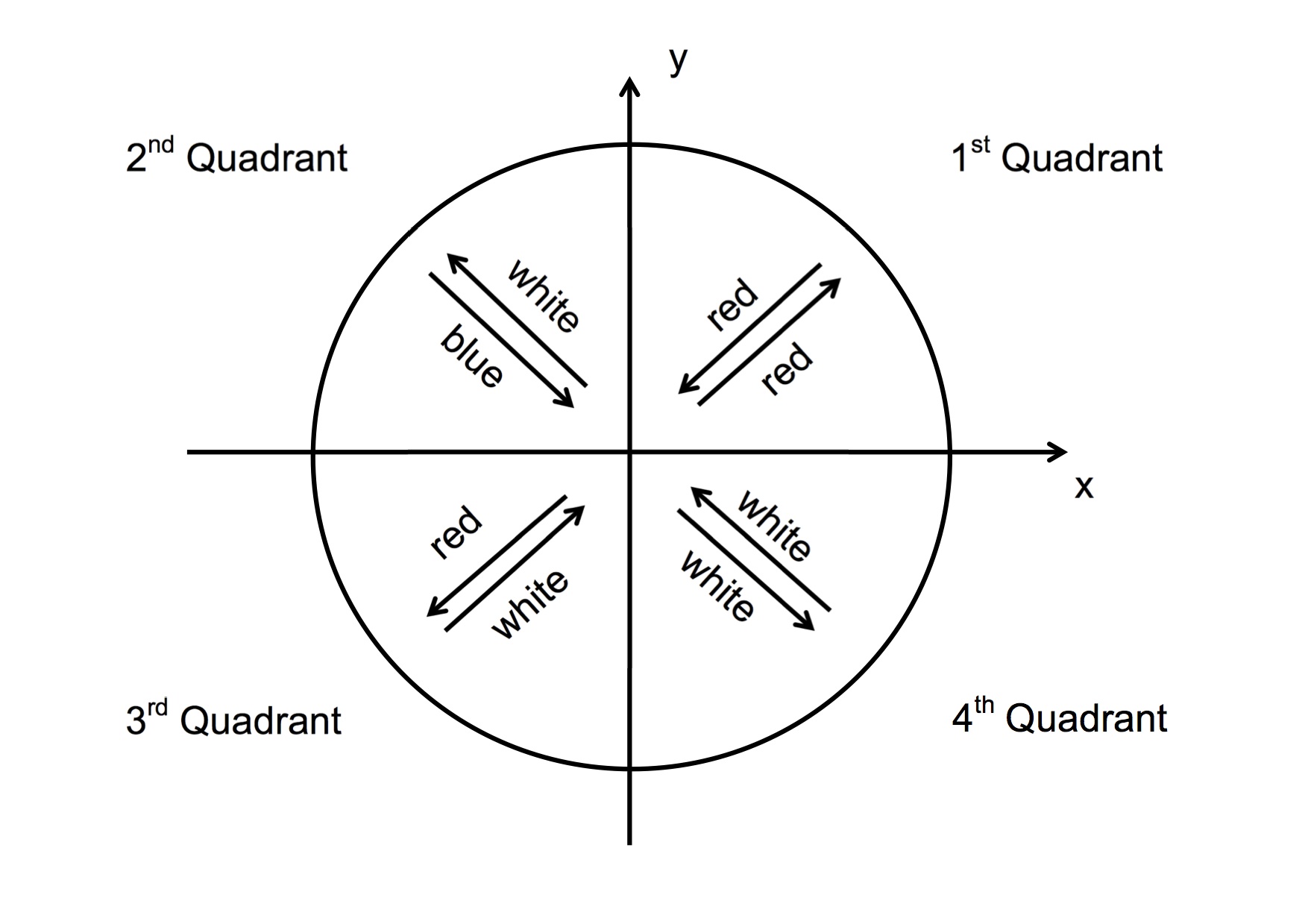}}\newline

\begin{center}
\small{Figure 4 --- The colour macro-variable $v''_{c}$.}
\end{center}
\end{figure}

Since the motion of the pendulum lies on a straight line through the midpoint, it always oscillates either between the first and the third quadrant, or between the second and the fourth quadrant. Therefore, for all trajectories with initial conditions either in the first or the third quadrant, the light is red 75\% of the time and white 25\% of the time; for trajectories with initial conditions in the second and the fourth quadrant the light is white 75\% of the time and blue 25\%  of the time. But neither white nor red is an equilibrium because half of all initial conditions lie on trajectories that only spend 25\% of the time in the white state, and the other half of initial conditions lie on trajectories that spend no time at all in the red state. This violates the requirement that initial conditions that don't spend most of the time in equilibrium form a set that has at most measure $\varepsilon \ll 1$.\footnote{This example also shows that the largest macro-state need not be the equilibrium state: $X''_{M_{w}}$ takes up $1/2$ of $X''$ and yet $M_{w}$ is not the equilibrium state.}\\

However, this seems to be the wrong conclusion because intuitively there are equilibria: for initial states with displacement coordinates in the first or the third quadrant the light is red 75\% of the time and hence red seems an equilibrium \emph{for states in those quadrants}, and likewise for initial states in the second or forth quadrant for which the light is white 75\% of the time. The root of the rift between mathematical criteria and intuition is that we tacitly took the entire state space $X''$ to be the effective state space $Z$, and with respect to $X''$ the conditions for the existence of an equilibrium are not satisfied. But nothing forces us to set $Z=X''$. In fact an alternative choice of $Z$ restores existence. Let $Z^{1}$ be the union of the first and the third quadrants. One can then easily construct the effective dynamical system  $(Z^{1},\Sigma_{Z^{1}},\mu_{Z^{1}},T'')$, where $\Sigma_{Z^{1}}$ is the Borel $\sigma$-algebra on $Z^{1}$, $\mu_{Z^{1}}$ is the measure $\mu_{X''}$ restricted to $Z^{1}$ and $T''$ is the dynamics restricted to $Z^{1}$. It is obvious that for that system the light being red is a an $0.75$-$0$-equilibrium of the $\alpha$-$\varepsilon$ type and a $0.5$-$0$-equilibrium of the $\gamma$-$\varepsilon$ type. And the same moves are available for the other two quadrants. Let $Z^{2}$ be the union of the second or the fourth quadrant. The corresponding effective dynamical system is $(Z^{2},\Sigma_{Z^{2}},\mu_{Z^{2}},T'')$, where $\Sigma_{Z^{2}}$ is the Borel $\sigma$-algebra on $Z^{2}$, $\mu_{Z^{2}}$ is the measure $\mu_{Z''}$ restricted to $Z^{2}$ and $T''$ is the dynamics restricted to $Z^{2}$. It is then obvious that the light being white is a an $0.75$-$0$-equilibrium of the $\alpha$-$\varepsilon$ type and a $0.5$-$0$-equilibrium of the $\gamma$-$\varepsilon$ type.\\

This example illustrates that the choice of the effective phase space $Z$ is crucial for the existence of an equilibrium. With the wrong choice of $Z$ -- the full three-dimensional state space -- no equilibrium exists. But if we choose either $Z^{1}$ or  $Z^{2}$ as the effective state space, then there are equilibria.\\

Let us explain why the Existence Theorem is satisfied for these effective dynamical systems. We first focus on $(Z^{1},\Sigma_{Z^{1}},\mu_{Z^{1}},T'')$. The index set is $\Omega''=\Omega_{3}\times\Omega_{4}$, where $\Omega_{3}$ consist of the possible energies of the system and each $\omega_{4}\in \Omega_{4}$, $\Omega_{4}=(0,\pi/2]$, denotes an angle and thus a line cutting through the coordinate origin in the first (and therefore also third) quadrant. The measure on $\Omega''$ arises from the product measure $\mu_{\Omega_{3}}\times \mu_{\Omega_{4}}$, where $\mu_{\Omega_{3}}$ is the uniform measure on the energy values and $\mu_{\Omega_{4}}$ is the uniform measure on $(0,\pi/2]$. Each $Z''_{\omega}$ is a two-dimensional ellipse determined by the initial energy and displacement coordinates; the sigma-algebra $\Sigma''_{\omega}$ is the usual Borel $\sigma$-algebra, and the measure $\mu''_{\omega}$ is the normalised length measure on the ellipse $Z''_{\omega}$. The dynamics on the ellipses is again given by the restriction of $T''_{t}$ to the ellipses $Z''_{\omega)}$. This gives us the components $C''_{\omega}=(Z''_{\omega},\Sigma_{\omega''},\mu''_{\omega}, T_{t}'')$. Again, it is clear that the motion on each component is ergodic. Now the Existence Theorem is satisfied for the same reason it is satisfied for the pendulum with a one-dimensional position coordinate, namely: equation (\ref{ggg}) holds true for every arbitrary component $C''_{\omega}$ because on each ellipse three-quarters of the states correspond to a red light and one quarter to a white light, and hence $\mu''_{\omega}(Z''_{\omega}\!\cap\! Z^{1}_{M_{r}}) \, \geq \, 0.75$. Similarly, equation (\ref{ggg})  holds true for every arbitrary component $C''_{\omega}$ because on each ellipse three-quarters of the states correspond to a white light and one quarter to a red light, and hence $\mu''_{\omega}(Z''_{\omega}\!\cap\! Z^{1}_{M_{r}}) \, \geq \, \mu''_{\omega}(Z''_{\omega}\!\cap\! Z^{1}_{M})+0.5$ for all $Z^{1}_{M}\neq Z^{1}_{M_{r}}$. Analogue reasoning for $Z^{2}$ shows that also for $(Z^{2},\Sigma_{Z^{2}},\mu_{Z^{2}},T'')$ the equations (\ref{ggg}) and (\ref{stress}) of the Existence Theorem are satisfied.\\

\section{A Fresh Look at the Ergodic Programme}\label{Ergodic}

The canonical explanation of equilibrium behaviour is given within the ergodic approach. Before looking at further examples, it is helpful to revisit this approach from the point of view of the Existence Theorem. We show that the standard ergodic approach in fact provides a triplet that satisfies the above conditions.\\

Many explanations of the approach to equilibrium rely on the dynamical conditions of ergodicity or epsilon-ergodicity (see Frigg 2008 and references therein). The definition of ergodicity was given above (Equation \ref{ergodicE}). A system $(Z,\Sigma_{Z},\mu_{Z},T_{t})$ is epsilon-ergodic iff it is ergodic on a set $\hat{Z}\subseteq Z$ of measure $1-\varepsilon$ where $\varepsilon$ is a very small real number.\footnote{In detail: $(Z,\Sigma_{Z},\mu_{Z},T_{t})$ is \textit{$\varepsilon$-ergodic}, $\varepsilon\in\field{R},\,0\leq\varepsilon<1$, iff there is a set $\hat{Z}\subset Z$, $\mu_{Z}(\hat{Z})=1-\varepsilon$, with $T_{t}(\hat{Z})\subseteq\hat{Z}$ for all $t$, such that the system
$(\hat{Z},\Sigma_{\hat{Z}},\mu_{\hat{Z}},T_{t})$ is ergodic, where $\Sigma_{\hat{Z}}$ and $\mu_{\hat{Z}}$ is the $\sigma$-algebra $\Sigma_{Z}$ and the measure $\mu_{Z}$ restricted to $\hat{Z}$. A system $(Z,\Sigma_{Z},\mu_{Z},T_{t})$
is \textit{epsilon-ergodic} iff there exists a very small $\varepsilon$ for which the system is $\varepsilon$-ergodic.}
The results of this paper clarify these claims. As pointed out in the previous subsection, if the macro-variables are not the right ones, then neither ergodicity nor epsilon-ergodicity imply that the approach to equilibrium takes place. However proponents of the ergodic approach often assume that there is a macro region which is either $\beta$-dominant or $\delta$-prevalent (e.g.\ Frigg and Werndl 2011, 2012). Then this leads to particularly simple instance the Existence Theorem, which then implies that the macro-region corresponds to an $\alpha$-$\varepsilon$-equilibrium or a $\gamma$-$\varepsilon$-equilibrium. More specifically, the following two corollaries hold (for proofs seee Werndl and Frigg 2015a):
\begin{quote}
\noindent\emph{Ergodicity-Corollary}:
 Suppose that the measure-preserving system $(Z,\Sigma_{Z},\mu_{Z},T_{t})$ is ergodic. Then the following are true:
(a) If the system has a macro-region $Z_{\hat{M}}$ that is $\beta$-dominant, $\hat{M}$ is an $\alpha$-$\varepsilon$-equilibrium for $\alpha=\beta$.
(b) If the system has a macro-region $Z_{\hat{M}}$ that is $\delta$-prevalent, $\hat{M}$ is a $\gamma$-$\varepsilon$-equilibrium for $\gamma=\delta$.
\end{quote}

\begin{quote}
\noindent\emph{Epsilon-Ergodicity-Corollary}:
\noindent Suppose that the measure-preserving system $(Z,\Sigma_{Z},\mu_{Z},T_{t})$ is epsilon-ergodic. Then the following are true:
(a) If the system has a macro-region $Z_{\hat{M}}$ that is $\beta$-dominant for $\beta-\varepsilon>\frac{1}{2}$, $Z_{\hat{M}}$ is a $\alpha$-$\varepsilon$-equilibrium for $\alpha=\beta-\varepsilon$.
(b) If the system has a macro-region $Z_{\hat{M}}$ that is $\delta$-prevalent for $\delta-\varepsilon>0$, $Z_{\hat{M}}$ is a $\gamma$-$\varepsilon$-equilibrium for $\gamma = \delta-\varepsilon$.
\end{quote}

It is important to keep in mind, however, that ergodicity and epsilon-ergodicity are just examples of dynamical conditions for which an equilibrium exists. As shown by the Existence Theorem, the dynamics need not be ergodic or epsilon-ergodic for there to be an equilibrium. \\


\section{Gases}\label{Gases}

We now discuss gas systems that illustrate the core theorems of this paper. We start with well-known examples -- the dilute gas, the ideal gas and the Kac gas -- and then turn to lesser-known systems that illustrate the role of the ergodic decomposition and the $\varepsilon$-set of initial conditions that can be excluded. We first discuss a simple example where the dynamics is ergodic, namely a gas of noninteracting particles in a stadium-shaped box. Then we turn to an example of a  system with an $\varepsilon$-set that is excluded because the system is epsilon-ergodic, namely a gas of noninteracting particles in a mushroom-shaped box. Finally, we examine a more complicated gas system where there are several ergodic components and an $\varepsilon$-set that is excluded, namely a gas of noninteracting particles in a multi-mushroom box.

\subsection{The Dilute Gas}

A dilute gas is a system a system of $N$ particles in a finite container isolated from the environment. Unlike the particles of the ideal gas (which we consider in the next subsection), the particles of the dilute gas do interact with each other, which will be important later on. We first briefly review the standard derivation of the Maxwell-Boltzmann distribution with the combinatorial argument and then explain how the argument is used in our framework.\\

A point $x=(q,p)$ in the $6N$-dimensional set of possible position and momentum coordinates $X$ specifies a \emph{micro-state} of the system. The classical Hamiltonian $H(x)$ determines the dynamics of the system. Since the energy is preserved, the motion is confined to the $6N-1$ dimensional energy hyper-surface $X_{E}$ defined by $H(x)=E$, where $E$ is the energy of the system. $X$ is endowed with the Lebesgue measure $\mu$, which is preserved under $T_{t}$. With help of $\mu$ a measure $\mu_{E}$ on $X_{E}$ can be defined which is preserved as well and is normalised, i.e.\  $\mu_{E}(X_{E})=1$ (\emph{cf.}\ Frigg 2008, 104).\\

To derive the Maxwell-Boltzmann distribution we consider the $6$-dimensional state space $X_{1}$ of one particle. The state of the entire gas is given by by $N$ point in $X_{1}$. Because the system has constant energy $E$ and is confined to a finite container, only a finite part of $X_{1}$ is accessible. This accessible part of $X_{1}$ is partitioned into cells of equal size $\delta^{dg}$ whose dividing lines run parallel to the position and momentum axes. This results in a finite partition $\Omega_{dg} := \{\omega^{dg}_{1}, \, ..., \omega^{dg}_{l}\}$, $l\in\field{N}$ (`dg' stands for `dilute gas'). The cell in which a particle's state lies is its \textit{coarse-grained micro-state}. An \emph{arrangement} is a specification of coarse-grained micro-state of each particle. Let $N_{i}$be the number of particles whose state is in cell $\omega^{dg}_{i}$. A \textit{distribution} $D=(N_{1}, N_{2},\ldots,N_{l})$ is a specification of the number of particles in each cell. Several arrangements are compatible with each distribution, and the number $G(D)$ of arrangements compatible with a given distribution $D$ is $G(D) = N!\, / \, N_1!N_2!\ldots,N_l!$. Boltzmann (1877) assumed that the energy $e_i$ of particle $i$ depends only on the cell in which it is located (and not on interactions with other particles), which allows him to express the total energy of the system as a sum of single particle energies: $E=\sum_{i=1}^{l}N_{i}e_{i}$. Assuming that the number of cells in $\Omega_{dg}$ is small compared to the number of particles, Boltzmann was able to show that $\mu_{E}(Z_{D_{dg}})$ is maximal if
\begin{equation}\label{rig}
N_i = B e^{\Delta e_{i}},
\end{equation}
where $B$ and $\Delta$ are parameters which depend on $N$ and $E$. This is the \emph{discrete Maxwell-Boltzmann distribution}, which we refer to as $D_{MB}$ \\

Textbooks wisdom has it that the Maxwell-Boltzmann distribution defines the equilibrium state of the gas. While not wrong, this is only part of a longer story. We have to introduce macro variables and define $Z$ before we can say what the system's macro-regions are, and only once these are defined we can check whether the dynamics is such that one of those macro-regions qualifies as the equilibrium region.\\

Let us begin with macro-variables. The macro-properties of a gas depend only on the distribution $D$. Let $W$ be a physical variable on the one-particle phase space. For simplicity we assume that this variable assumes constant values $w_{j}$ in cell $\omega^{dg}_{j}$ for all $j=1, ..., l$. Physical observables can then written as averages of the form $\sum_{j=1}^{N} w_{j} N_{j}$ (for details see Tolmeman 1938, Ch. 4). It is obvious that every point $x \in X$ is associated with exactly one distribution $D$, which we call $D(x)$. Given $D(x)$ one can calculate $\sum_{j=1}^{N} w_{j} N_{j}$ at point $x$, which assigns every point $x$ a unique value. Hence a physical variable $W$ and a distribution $D(x)$ induce a mapping from $X$ to a set of values. Let us call this mapping $v$, and so we can write:   $v:X\rightarrow \field{V}$, where $\field{V}$ is the range of certain physical variable. Choosing different $W$ (with different $w_{j}$) will lead to different a different $v$. These are the macro-variables of the kind introduced in Section \ref{BSM}. A set of values of these variables defines a \emph{macro-state}. For the sake of simplicity we now assume that this set of values would be different for every distribution so that there is a one-to-one correspondence between distributions and macro-states.\\

The Maxwell-Boltzmann distribution depends on the total energy of the system: different energies lead to different equilibrium distributions. This tells us that equilibrium has to be defined with respect to the energy hyper-surface $X_{E}$.\footnote{Note that this is one of crucial differences between the dilute gas and the oscillator with a colour macro-variable of Section \ref{Simple Example}: the colour equilibrium does not depend on the system's energy.} States of different energy can never evolve into the same equilibrium and therefore no equilibrium state exists with respect to the full state space $X$. Now the assumption that the particles of the dilute gas interact becomes crucial. If the particles did not interact, there could be constants of motion other than the total energy and this might have the consequence equilibrium would have to be defined on a subsets of $X_{E}$ (we discuss such a case in the next subsection). It is usually assumed that this is not the case. The effective state space $Z$ then is $X_{E}$, and $(X_{E},\Sigma_{E},\mu_{E},T_{t})$ is the effective measure-preserving dynamical system of the dilute gas, where $\Sigma_{E}$ is the the Borel $\sigma$-algebra of $X_{E}$ and $T_{t}$ is the flow of the system restricted to  $X_{E}$. \\

We can now construct the macro-regions $Z_{M}$.  Above we assumed that there is a one to one correspondence between distributions and macro-states. So let $M_{D}$ be the macro state corresponding to distribution $D$. The macro-region $Z_{M_{D}}$ is then just the set of all $x \in X_{E}$ that are associated with $D$: $Z_{M_{D}} = \{x\in X_E\,\,:\,\,D(x)=D\}$. A fortiori this also provides a definition of the macro-state $M_{D_{MB}}$ associated with the Maxwell-Boltzmann distribution $D_{MB}$. Let us call the macro-region associated with that macro-state $Z_{MB}$. It is generally assumed that $Z_{MB}$ is the largest of all macro-states (relative to $\mu_{E}$), and we follow this assumption here.\footnote{The issue is the following. Equation $(\ref{rig})$ gives the distribution of largest size relative to the Lebesgue measure on the $6N$-dimensional shell-like domain $X_{ES}$ specified by the condition that $E=\sum_{i=1}^{l}N_{i}e_{i}$. It does \emph{not} give us the distribution with the largest measure $\mu_{E}$ on the $6N-1$ dimensional $Z_{E}$. Strictly speaking nothing about the size of $Z_{MB}$ (with respect to $\mu_{E}$) follows from the combinatorial considerations leading to Equation $(\ref{rig})$. Yet it is generally assumed that the proportion of the areas corresponding to different distributions are the same on  $X$  and  on $X_{E}$ (or at least that the relative ordering is the same). Under that assumption $Z_{E}$ is indeed the largest macro-region. We agree with Ehrenfest and Ehrenfest (1959, 30) that this assumption is in need of further justification, but grant it for the sake of the argument.}\\

Even if we grant that $Z_{MB}$ is the largest macro-region (in one of the senses of `large'), it is not yet clear that $M_{D_{MB}}$ is the equilibrium macro-state (in one of the senses of `equilibrium'). It could be that the dynamics is such that initial conditions that lie outside $Z_{MB}$ avoid $Z_{MB}$, or that a significant portion of initial conditions lie on trajectories that spend only a short time in $Z_{MB}$. To rule out such possibilities one has to look at the dynamics of $T_{t}$. Unfortunately the dynamics of dilute gases is mathematically not well understood, and there is no rigorous proof that the dynamics is `benign' (meaning that it does not have any of the features just mentioned). However, there are plausibility arguments for the conclusion that $T_{t}$ is epsilon-ergodic (Frigg and Werndl 2011). If these arguments are correct, then the dilute gas falls under the Epsilon-Erogidicity-Corollary and $Z_{MB}$ is an equilibrium either of the $\alpha$-$\varepsilon$ or the $\gamma$-$\varepsilon$ type, depending on whether $Z_{MB}$ is $\beta$-dominant or $\delta$-prevalent. Moreover, even if the dynamics turned out not to be epsilon-ergodic, it is a plausible assumption that the dynamics is such that the conditions of the Existence Theorem is fulfilled.\\

Hence Maxwell-Boltzmann distribution corresponds to equilibrium as expected. However, the above discussion shows that this does not come for free: we have to accept that $Z_{MB}$ is large and that $T_{t}$ is epsilon-ergodic, and making these assumptions plausible the choice of the right effective state space $Z$ is crucial. In fact, \emph{relative to $X$ no equilibrium exists} because there are different equilibria for different total energies of the system (as reflected by the Maxwell-Boltzmann distribution, which depends on the total energy $E$). This shows that the triplet of macro-variables, dynamics, and effective state space has to be well-adjusted for an equilibrium to exists, and that even small changes in one component can destroy this balance. \\

\subsection{The Ideal Gas}\label{ideal gas}
Now consider an ideal gas, a system consisting of $N$ particles with mass $m$ and \emph{no interaction at all}. We consider the same partitioning of the phase space as above and hence can consider the same distributions and the same macro-variables. One might then think that the ideal gas is sufficiently similar to the dilute gas to regard $Z_{MB}$ as the equilibrium state and lay the case to rest.\\

This is a mistake. To see why we need to say more about the dynamics of the system. An common way to describe the gas mathematically is to assume that the particles move on a three three-dimensional torus with constant momenta in each direction.\footnote{One also think of the particles as bouncing back and forth in box. In this case the modulo of the momenta is preserved and a similar argument applies.} This implies that  \emph{all} one-particle particle momenta $p_i$ (and hence all one-particle energies $e_i=p_i^{2}/2m$) are conserved quantities. As a consequence, if an ideal gas starts in a micro-state in which the momenta of the particles are not distributed according to the Maxwell-Boltzmann distribution, they will never reach that distribution. In fact the initial distribution is preserved no matter what that distribution is. For this reason $Z_{MB}$ is not the equilibrium state and the Maxwell-Boltzmann distribution does \emph{not} characterise the equilibrium state. So the combinatorial argument does not provide the correct equilibrium state for an ideal gas, and $Z_{E}$ is not the effective state space.\\

This, however, does not imply that the ideal gas has no equilibrium at all. Intuitively speaking, there is a $\gamma$-$\varepsilon$-equilibrium, namely the one where all particles are uniformly distributed. To make this more explicit let us separate the distribution $D$ into the position distribution $D_{x}$ and the momentum $D_{p}$ (which is a trivial decomposition which can always be done): $D= (D_{x}$, $D_{p}$). Under the dynamics of the ideal gas $D_{p}$ will not change over time and hence remain in whatever initial distribution the gas is prepared. By contrast, the position distribution $D_{x}$ will approach an even distribution $D_{e}$ as time goes on. So we can say that the equilibrium distribution of the system is $D_{eq}= (D_{e}$, $D_{p}$), where $D_{p}$ is the gas' initial distribution. The relevant space with respect to which an equilibrium exists is the hyper-surface $Z_{p}$, i.e. the hyper-surfaface defined by the condition that the moment are distributed according to $D_{p}$. The relevant dynamical system then is $(Z_{p}, \Sigma_{Z_{p}},\mu_{p},T_{t})$, where $\Sigma_{Z_{p}}$ is the Borel-$\sigma$-algebra, $\mu_{p}$ is the uniform measure on $Z_{p}$ and the dynamics $T_{t}$ is simply the dynamics of the ideal gas restricted to $Z_{p}$.\\

It is easy to see that with respect to $Z_{p}$ the region corresponding to $D_{e}$ is the largest macro-region. The motion on $Z_{p}$ is  ergodic for almost all momentum coordinates.\footnote{In terms of the uniform measure on the momentum coordinates.} Thus, by the Ergodicity Corollary, the largest macro-region, i.e. the macro-region corresponding to the uniform distribution, is a $\gamma$-$\varepsilon$-equilibrium. There will be some very special momentum coordinates where no equilibrium exists relative to $Z_{p}$ because the motion of the particles is periodic. However, these special momentum coordinates are of measure zero and for all other momentum coordinates the uniform distribution will correspond to the equilibrium macro-region. Thus this example illustrates again the importance of choosing the correct effective phase space: the ideal gas has no equilibrium relative to $\Gamma_{E}$ but an equilibrium exists relative to $Z_{p}$.\footnote{Another possible treatment of the ideal gas is to consider the different macro-state structure given only by the coarse-grained position coordinates (i.e.\ the momentum coordinates are not considered). Then the effective dynamical system would coincide with the full dynamical system $(\Gamma,\Sigma_{\Gamma},\mu_{\Gamma},T_{t})$. Relative to this dynamical system there would be an $\gamma$-$0$-equilibrium (namely the uniform distribution). That is, almost all initial conditions would spend most of the time in the macro-state that corresponds to the uniform distribution of the position coordinates.}

\subsection{The Kac Gas}\label{KacS}

The Kac-ring model consists of an even number $N$ of sites distributed equidistantly around a circle. On each site there is either a white or black ball. Let us assume that $N/2$ of the points (forming a set $S$) between the sites of the balls are marked. A specific combination of white and black balls for all sites together with the set $S$ is a \emph{micro-state} $k$ of the system, and the state space $K$ consists of all combinations of white and black balls and selection of $N/2$ points between the sites and $\Sigma_{K}$ is the power set of $K$. The dynamics $\kappa$ of the system is given as follows: during one time step each ball moves counterclockwise to the next site; if the ball crosses an interval in $S$, it changes colour and if it does not cross an interval in $S$, then it does not change colour (the set $S$ stays the same at all times). The probability measure is the uniform measure $\mu_{K}$ on $K$. $(K, \Sigma_{K},\mu_{K}, \kappa_{t})$, where $\kappa_{t}$ is the $t$-th iterate of $\kappa$ is a measure-preserving deterministic system describing the behaviour of the balls (and $K$ is both the full state space $X$ as well as the effective state space $Z$ of the system). \\

The Kac-ring can be interpreted in several ways. As presented here, the intended interpretation is that of a gas: the balls are described by their positions and their colour is seen as representing their (discrete) velocity. Whenever a ball passes a marked site its colour changes, which is analogous to a change in velocity of a molecule that results from collision with another molecule. The equations of motion are given by the counterclockwise motion together with the changing of the colours (Bricmont 1995; Kac 1959; Thompson 1972). The macro-states usually considered are defined by the total number of black and white balls. So the relevant macro-varible $v$ is a mapping
$K\rightarrow \field{V}$, where $\field{V}=\{0, ..., N\}$. Each value in $\field{V}$ defines a different macro-state. Traditionally these states are labelled $M^{K}_{i}$, where $i$ denotes the total number of white balls, $0\leq i\leq N$. As above, the \emph{macro-regions} $K_{i}$ are defined as the set of micro-states on which $M^{K}_{i}$ supervenes. It can be shown that the macro-state whose macro-region is of largest size is $M^{K}_{N/2}$, i.e.\ the state in which half of the spins are up and half down.\\

This example is interesting because it illustrates the case where an equilibrium exists even though the phase space is broken up into a finite number of ergodic components. More specifically, the motion of the Kac-ring is periodic. Suppose that $N/2$ is even: then at most after $N$ steps all balls have returned to their original colour because each interval has been crossed once and $N/2$ is even. If $N/2$ is odd, then it takes at most $2N$ steps for the balls to return to their original colour (because after $2N$ steps each interval has been crossed twice). So the phase space of the KAC-ring is decomposed into periodic cycles (together with a specification of $S$). These cycles are the components of the ergodic decomposition that we encounter in the Ergodic Decomposition Theorem. The Existence Theorem is satisfied and hence a $\gamma$-$\varepsilon$-equilibrium exists because on each of these ergodic components, except for components of measures $\varepsilon$, the equilibrium macro-state $M^{K}_{N/2}$ takes up the largest measure, i.e.\ Equation (\ref{stress}) is satisfied. Note that there are initial states that do not show equilibrium-like behaviour (that is, the set of initial conditions that do not show an approach to equilibrium is of positive measure $\varepsilon$). For instance, start with all balls being white and let every interval belong to $S$. Then, clearly, after one step the balls are all black, then after one step they are all white, and so on there is no approach equilibrium (Bricmont 2001; Kac 1959; Thompson 1972). The Existence Theorem is satisfied and hence a $\gamma$-$\varepsilon$-equilibrium exists because on each of these ergodic components, except for components of measures $\varepsilon$, the equilibrium macro-state $M^{K}_{N/2}$

\subsection{Gas of Noninteracting Particles in a Stadium-Box}

Let us now turn to lesser-known examples of gas systems that illustrate the various cases of the Existence Theorem. The first example illustrates the easiest way to satisfy the existence theorem, namely having an ergodic dynamics and a macro-region of largest measure. Consider a stadium-shaped box $S$ (i.e.\ a rectangle capped by semicircles). Suppose that $N$ particles are moving with uniform speed\footnote{Speed, unlike velocity, is not directional and does not change when particle bounces off the wall.} inside the stadium-shaped box, where the collisions with the walls are assumed to be elastic and it is further assumed that the particles do not interact. The set of all possible states of the system consists of the points $Y=(y_{1},w_{1},y_{2},w_{2}\ldots,y_{N},w_{N})$ satisfying the constraints $y_{i}\in S$ and $||w_{i}||=1$, where $y_{i}$ and $w_{i}$ are the position and velocity coordinates of the particles respectively $(1\leq i \leq N)$. $\Sigma_{Y}$ is the Borel $\sigma$-algebra of $Y$. The dynamics $R_{t}$ of the system is the motion resulting from particles bouncing off the wall elastically (whithout interacting with each other). The uniform measure $\nu$ is the invariant measure of the system. $(Y, \Sigma_{Y}, \nu, R_{t})$ is a measure-preserving dynamical system and it can be proven that the system is ergodic (\emph{cf.} Bunimovich 1979).\footnote{Bunimovich's (1979) results are about \emph{one} particle moving in a stadium-shaped box, but they immediately imply the results stated here about $n$ non-interacting particles.} $Y$ is both the full state space $X$ and the effective state space $Z$ of the system. \\

Now divide the stadium-shaped box into cells $\omega^{S}_{1}, \omega^{S}_{2}, \ldots,\omega^{S}_{l}$ of equal measure $\delta^{S}$ ($l\in\field{N}$). As in the case of the dilute gas, consider distributions $D=(N_{1}, ..., N_{l})$ and associate macro-states with these distributions. Macro-variables are also defined as above. It is then obvious that the macrostate $(N/l, N/l,\ldots, N/l)$ corresponds to the macro-region of largest measure.\footnote{It is assumed here that $N$ is a multiple of $l$.} Since the dynamics is ergodic, it follows from the Ergodicity-Corollary that the system has a $\gamma$-$\varepsilon$-equilibrium (where $\varepsilon=0$). More specifically, except for a set of measure zero, for all initial states of the $N$ billiard balls the system will approach equilibrium and stay there most of the time.

\subsection{Gas of Noninteracting Particles in a Mushroom-Box}
The next example illustrates the role of the $\varepsilon$-set of initial conditions that are not required to show equilibrium-like behaviour in the Definition of a $\gamma$-$\varepsilon$-equilibrium. For most conservative systems the phase space is expected to consist of regions of chaotic or ergodic behaviour next to regions of regular and integrable behaviour. These mixed systems are notoriously difficult to study analytically as well as numerically (Porter and Lansel 2006). So it was a considerable breakthrough when Bunimovich (2002) introduced a class of billiard systems that can easily be shown to have mixed behaviour.\\

\begin{figure}
\centering
\resizebox{0.7\textwidth}{!}{\includegraphics{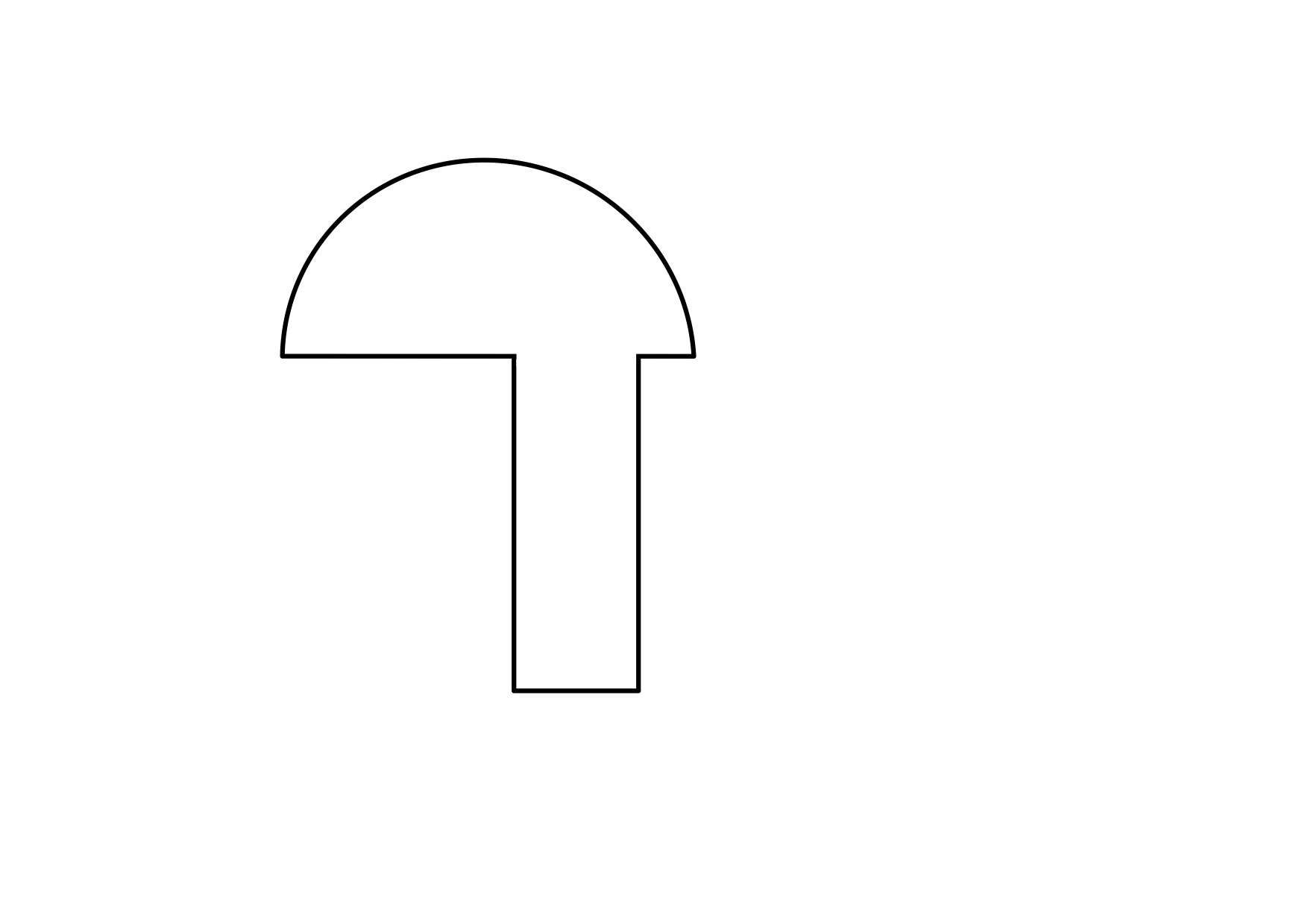}}

\begin{center}
\small{Figure 5 --- The mushroom-shaped box.}
\end{center}
\end{figure}

Consider a mushroom-shaped box (the domain $M$ obtained by placing an ellipse on top of a rectangle as shown in Figure 5), consisting of the stem $St$ and the cap $Ca$.  Suppose that $N$ gas particles are moving with uniform speed inside the mushroom-shaped box. The collisions on the wall are again assumed to be elastic and, for sake of simplicity, we assume that the particles do not interact. Then the set of all possible states consists of the points $D=(d_{1},v_{1},d_{2},v_{2}\ldots,d_{n},v_{n})$, where $d_{i}\in M$ and $||v_{i}||=1$ are the position and velocity coordinates of the particles respectively $(1\leq i \leq N)$. $\Sigma_{D}$ is the Borel $\sigma$-algebra of $D$. The dynamics $U_{t}$ of the system is the motion of the particles generated by elastic collisions with the boundaries of the mushroom. The phase volume $u$ is preserved under the dynamics. $(M, \Sigma_{M}, U_{t}, u)$ is a measure-preserving dynamical system (and $M$ is both the full state space $X$ and the effective state space $Z$ of the system). \\

It can be proven that the phase space consist of two regions: an ergodic region and a region with regular or mixed behaviour (i.e. integrable parts are intertwined with chaotic parts). As the stem is shifted to the left, the volume of phase space occupied by the ergodic motion continually increases and finally reaching measure 1 when the stem reaches the edge of the cap.\footnote{The results in Bunimovich (2002) are all about \emph{one} particle moving inside a mushroom-shaped box, but they immediately imply the results about a system of $n$ non-interacting particles stated here.} Assume now that the stem be so far to the left that
the measure of the ergodic region is $1-\varepsilon$, in which case the system is $\varepsilon$-ergodic (\emph{cf.}  Section \ref{Ergodic}). Suppose that the macro-states of interest are distributions $D=(N_{St}, N_{Ca})$, where $N_{St}$ and $N_{Ca}$ are the particle numbers in the stem and cap respectively. We now assume that that the measure of the stem is the same as the measure of the cap. It then follows from the Epsilon-Ergodicity Corollary that the system has a $\gamma$-$\varepsilon$ equilibrium, namely $(N/2, N/2)$ (\emph{cf.} Bunimovich 2002; Porter and Lansel 2006). This example is of special interest because it is proven that there is a set of initial states of the billiard balls of positive measure that do not show equilibrium-like behaviour. That is, for these initial states the system does not evolve in such a way that most of the time half of the particles are in the stem and half of the ball are in the cap (as is allowed by the definition of an $\gamma$-$\varepsilon$-equilibrium).

\subsection{Gas of Noninteracting Particles in a Multi-Mushroom-Box}
In our next and last example the Existence Theorem is satisfied because there are \emph{a finite number of ergodic components} on each of which the equilibrium macro-state takes up the largest measure. Consider a box created by several mushrooms such as the one shown in Figure 6 (the domain $TM$ is constructed from three elliptic mushrooms, where the semi-ellipses have foci $F_1$ and $F_2$, $F_3$ and $F_4$, $F_5$ and $F_6$).\\

\begin{figure}
\centering
\resizebox{0.7\textwidth}{!}{\includegraphics{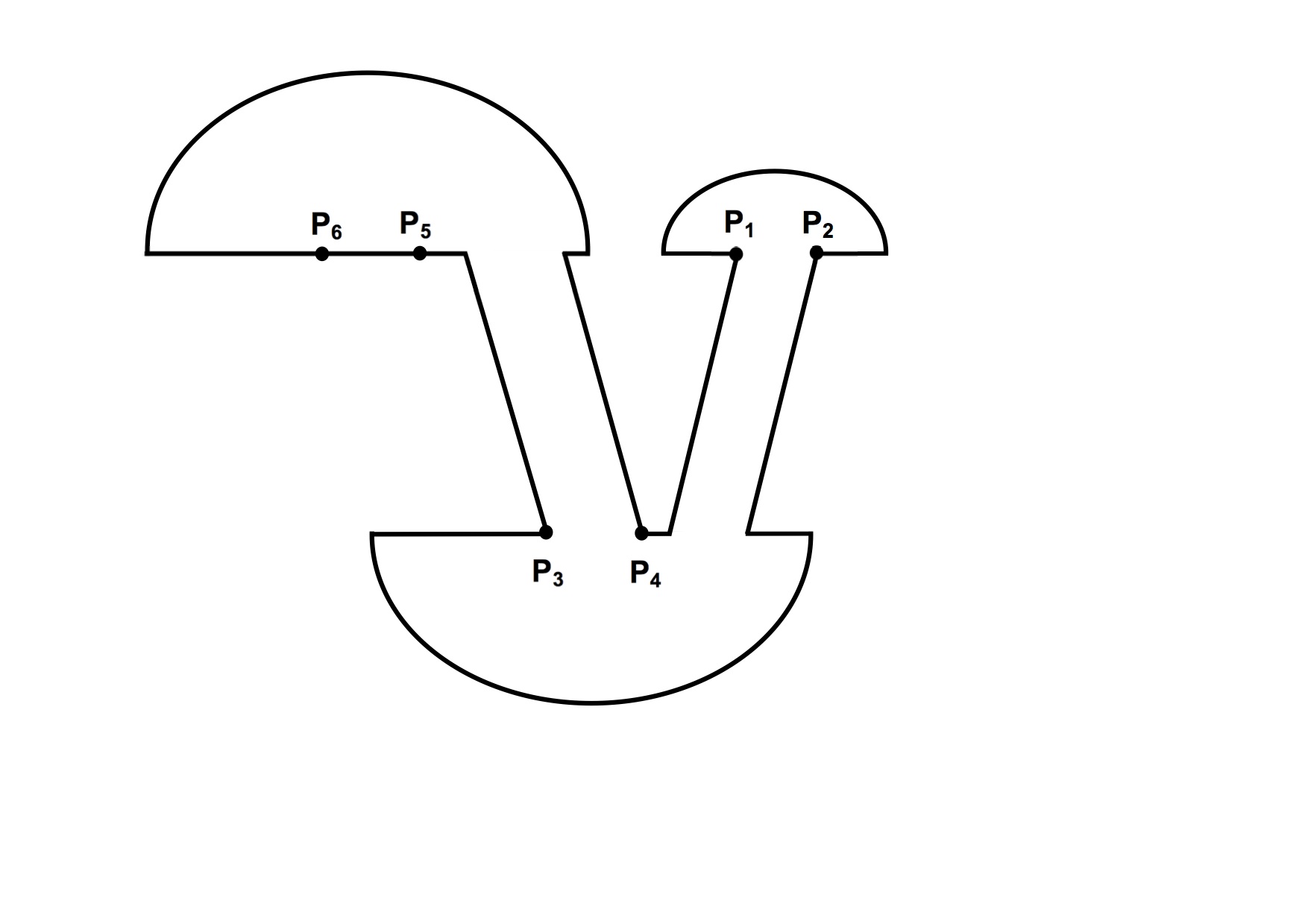}}

\begin{center}
\small{Figure 6 --- The multi-mushroom-shaped box.}
\end{center}
\end{figure}

Suppose again that $N$ gas particles are moving with uniform speed inside the box $MM$, that the collisions on the wall are elastic and that the particles do not interact. The set of all possible states consists of the points $W=(w_{1},u_{1},w_{2},u_{2}\ldots,w_{n},u_{n})$, where $w_{i}\in MM$ and $||u_{i}||=1$ are the position and velocity coordinates of the particles respectively $(1\leq i \leq N)$. $\Sigma_{W}$ is the Borel $\sigma$-algebra on $W$. The dynamics $V_{t}$ of the system is given by the motion of the noninteracting particles inside the box, and the phase volume $v$ is preserved under the dynamics. $(W, \Sigma_{W}, v, V_{t})$ is a measure-preserving dynamical system (and $W$ is both the full state space $X$ and the effective state space $Z$ of the system).\\

Bunimovich (2002) proved that the phase space consist of $2^{N}$ larger regions on each of which the motion is ergodic and, finally, one region of negligible measure $\varepsilon$ of regular or mixed behaviour (Bunimovich 2002).\footnote{How small $\varepsilon$ is depends on the exact shape of the box of the three elliptic mushrooms (it can be made arbitrarily small).} The $2^{N}$ ergodic components arise in the following way: each single particle space has two large ergodic regions: one region consisting of those orbits of the particle that move back and forth between the semi-ellipses with the foci $P_1$, $P_2$ and $P_3$, $P_4$ (while never visiting the semi-ellipse with the foci $P_5, P_6$). The second  ergodic region consists of the orbits that travel back and forth between the semi-ellipses with the foci $P_3$, $P_4$ and $P_5$, $P_6$ (while never visiting the semi-ellipse with the foci $P_1, P_2$). Given that the phase space of the entire system is just the cross-product of the phase space of the $N$ single particle spaces, it follows that there are $2^{N}$ ergodic components.\footnote{Again, Bunimovich's (2002) results are all about \emph{one} particle moving inside a mushroom-shaped box, but they immediately imply the results about a system of $n$ non-interacting particles stated here.} \\

Suppose that the macrostates of interest are the distributions $D=(MS,MC)$, where $MS$ is the number of balls in the two stems and $MC$ the number of balls in the three caps of the mushroom, where we assume that the measure of the two stems taken together is the same as the measure of the three caps taken together.\footnote{This can always be arranged in this way -- see Bunimovich (2002).} Then the system has a $\gamma$-$\varepsilon$ equilibrium, corresponding to the case where $N/2$ of the particles are in the two stems and $N/2$ of the particles are in the three caps of the mushrooms (cf.\ Bunimovich 2002; Porter and Lansel 2006). This example is of special interest because it illustrates the case where an equilibrium exists even though the phase space is broken up into a finite number of ergodic components. More specifically, in this case we encounter $2^{N}$ ergodic components on each of which the equilibrium macro-state takes up the largest measure and hence equation (\ref{stress}) is satisfied. Since these ergodic components taken together have total measure $1-\varepsilon$ and the definition of a $\gamma$-$\varepsilon$-equilibrium allows that there is an $\varepsilon$ set of initial conditions that do not show equilibrium-like behaviour, it follows that the Existence Theorem is satisfied.

\section{Conclusion}\label{Conclusion}

In this paper we introduced a new definition of Boltzmannian equilibrium and presented an existence theorem that characterises the circumstances under which a Boltzmannian equilibrium exists. The definition and the theorem are completely general in that they make no assumption about the nature of interactions and so they provide a characterisation of equilibrium also in the case of strongly interacting systems. The approach also ties in smoothly with the Generalised Nagel-Schaffner model of reduction (Dizadji-Bahmani \emph{et al.} 2010) and hence serves as a starting point for discussions about the reduction of thermodynamics to statistical mechanics.
\\

The framework raises a number of questions for future research. First, our discussion is couched in terms of deterministic dynamical systems. In a recent paper (Werndl and Frigg 2016) we generalise the definition of equilibrium to stochastic systems. To date there is, however, no such generalisation of the Existence Theorem. The reason for this is that this theorem is based on the ergodic decomposition theorem, which has no straightforward stochastic analogue. So it remains an open question how circumstances under which an equilibrium exists in stochastic systems should be characterised.\\

Second, macro-variables raise a number of interesting issues. An important question is: how exactly do the macro-variables look like for the variety of physical systems discussed in statistical  mechanics? It has been pointed out to us\footnote{By David Lavis and Reimer K\"uhn in private conversation.} that for intensive variables the exact definition is complicated and can only be done by referring to extensive quantities. A further issue is that many quantities of interest are \emph{local quantities}, at least as long as the system is not in equilibrium (pressure and temperature are cases in point). Such quantities have to be described as fields, which requires an extension of the definition of a macro-state in Section \ref{BSM}. Rather than associating equilibrium only with a certain value (or a range of values), one now also has to take field properties such as homogeneity into account.\\

Finally, there is a question about how to extend our notion of equilibrium to quantum systems. Noting in our definition of equilibrium depends on the underlying dynamics being classical or the variables being defined on a classical space phase space rather than a Hilbert space, and so we think that there are no in-principle obstacles to carrying over our definition of equilibrium to quantum mechanics. But the proof of the pudding is in the eating and so the challenge is to give an explicit quantum mechanical formulation of Boltzmannian equilibrium.

\section*{References}

\begin{list}{}{    \setlength{\labelwidth}{0pt}
    \setlength{\labelsep}{0pt}
    \setlength{\leftmargin}{24pt}
    \setlength{\itemindent}{-24pt}
  }

\item Bricmont, J. (2001). Bayes, Boltzmann and Bohm: Probabilities in Physics. In \emph{Chance in Physics: Foundations and Perspectives}, ed. J. Bricmont, D. D\"{u}rr, M.C. Galavotti, G.C. Ghirardi, F. Pettrucione, and N. Zanghi. Berlin and New York: Springer, 3--21.

\item Bunimovich, L.A. (1979). On the Ergodic Properties of Nowhere Dispersing Billiards. \emph{Communications in Mathematical Physics} 65, 295-312.

\item Bunimovich, L.A (2002). Mushrooms and Other Billiards With Divided Phase Space. \emph{Chaos} 11 (4), 802-808.

\item Callender, C. 2001. Taking Thermodynamics Too Seriously. \emph{Studies in History and Philosophy of Modern Physics} 32, 539-553.

\item Dizadji-Bahmani., F., Frigg, R. and S. Hartmann (2010). Who's Afraid of Nagelian Reduction? \emph{Erkenntnis} 73, 393-412.

\item Ehrenfest, P. and Ehrenfest, T. (1959). \emph{The Conceptual Foundations of the Statistical Approach in Mechanics}. Ithaca, New York: Cornell University Press.

\item Frigg, R. and Werndl, C. (2011). Explaining Thermodynamic-Like Behaviour in Terms of Epsilon-Ergodicity. \emph{Philosophy of Science} 78, 628-652.

\item Frigg, R. and Werndl, C. (2012). A New Approach to the Approach to Equilibrium.  In \emph{Probability in Physics}, ed. Y. Ben-Menahem and M. Hemmo. The Frontiers Collection. Berlin: Springer, 99-114.

\item Kac, M. (1959). \emph{Probability and Related Topics in the Physical Sciences}. New York: Interscience Publishing.

\item Petersen, K. 1983. \emph{Ergodic theory}. Cambridge: Cambridge University Press.

\item Porter, M.A. and Lansel, S. (2006). Mushroom Billiards. \emph{Notices of the American Mathematical Society} 53 (3), 334-337.

\item Reiss, H. 1996. \emph{Methods of Thermodynamics}. Mineaola/NY: Dover.

\item Sklar, L. (1973). \emph{Philosophical Issues in the Foundations of Statistical Mechanics}. Cambridge: Cambridge University Press.

\item  Thompson, C. J. (1972). \emph{Mathematical Statistical Mechanics}. Princeton: Princeton University Press.

\item Tolman,  R. C.  (1938).  \emph{The Principles  of  Statistical  Mechanics}.  Mineola/New York: Dover 1979.

\item Werndl, C. and Frigg, R. (2015a). Reconceptualising Equilibrium in
Boltzmannian Statistical Mechanics and Characterising its Existence. \emph{Studies in
History and Philosophy of Modern Physics} 44 (4), 470-479.

\item Werndl, C. and Frigg, R. (2015b). Rethinking Boltzmannian Equilibrium. \textit{Philosophy of Science} 82 (5), 1224-1235.

\item Werndl, C. and Frigg, R. (2016). Boltzmannian Equilibrium in Stochastic Systems. Forthcoming in \emph{Proceedings of the European Philosophy of Science Association} 82 (5), 1224-1235.
\end{list}

\end{document}